\documentstyle[12pt,epsfig]{article}

\newskip\humongous \humongous=0pt plus 1000pt minus 1000pt

\newif\ifdtup

\def\pr#1{#1^\prime}

\def\beq{\begin{equation}}
\def\eeq{\end{equation}}

\def\beqn{\begin{eqnarray}}
\def\eeqn{\end{eqnarray}}
\relax

\def\dotx{\dotx{\dot\overline{x}}}

\relax

\catcode`\@=11

\@addtoreset{equation}{section}
\def\theequation{\thesection\arabic{equation}}

\def\@normalsize{\@setsize\normalsize{15pt}\xiipt\@xiipt
\abovedisplayskip 14pt plus3pt minus3pt%
\belowdisplayskip \abovedisplayskip
\abovedisplayshortskip \z@ plus3pt%
\belowdisplayshortskip 7pt plus3.5pt minus0pt}

\def\small{\@setsize\small{13.6pt}\xipt\@xipt
\abovedisplayskip 13pt plus3pt minus3pt%
\belowdisplayskip \abovedisplayskip
\abovedisplayshortskip \z@ plus3pt%
\belowdisplayshortskip 7pt plus3.5pt minus0pt
\def\@listi{\parsep 4.5pt plus 2pt minus 1pt
     \itemsep \parsep
     \topsep 9pt plus 3pt minus 3pt}}

\@twosidetrue

\relax

\catcode`@=12

\catcode`\@=11

\def\section{\@startsection{section}{1}{\z@}{3.5ex plus 1ex minus
   .2ex}{2.3ex plus .2ex}{\large\bf}}

\def\thesection{\arabic{section}.}

\def\appendix{\setcounter{section}{0}
 \def\thesection{APPENDIX \Alph{section}:}
 \def\theequation{\Alph{section}.\arabic{equation}}}

\def\ps@headings{\def\@oddfoot{}\def\@evenfoot{}
\def\@oddhead{\hbox{}\hfill
 \makebox[.5\textwidth]{\raggedright\ignorespaces --\thepage{}--
 \hfill {}}}  
\def\@evenhead{\@oddhead}
\def\subsectionmark##1{\markboth{##1}{}}
}

\ps@headings

\catcode`\@=12

\def\figcap{\section*{Figure Captions\markboth
 {FIGURECAPTIONS}{FIGURECAPTIONS}}\list
 {Fig. \arabic{enumi}:\hfill}{\settowidth\labelwidth{Fig. 999:}
 \leftmargin\labelwidth
 \advance\leftmargin\labelsep\usecounter{enumi}}}
 \relax
\def\tablecap{\section*{Table Captions\markboth
 {TABLECAPTIONS}{TABLECAPTIONS}}\list
 {Table \arabic{enumi}:\hfill}{\settowidth\labelwidth{Table 999:}
 \leftmargin\labelwidth
 \advance\leftmargin\labelsep\usecounter{enumi}}}
 \relax
\def\reflist{\section*{References\markboth
 {REFLIST}{REFLIST}}\list
 {[\arabic{enumi}]\hfill}{\settowidth\labelwidth{[999]}
 \leftmargin\labelwidth
 \advance\leftmargin\labelsep\usecounter{enumi}}}
 \relax

\catcode`\@=11

\def\ps@headings{\def\@oddfoot{}\def\@evenfoot{}
\def\@oddhead{\hbox{}\hfill
 \makebox[.5\textwidth]{\raggedright\ignorespaces --\thepage{}--
 \hfill {}}}    
\def\@evenhead{\@oddhead}
\def\subsectionmark##1{\markboth{##1}{}}
}

\ps@headings

\catcode`\@=12

\relax

\catcode`\@=11
\def\prm{\fam \z@}
\catcode`\@=12

\relax
\def\pl#1#2#3{{\it Phys. Lett. }{\bf #1}(19#2)#3}
\def\zp#1#2#3{{\it Z. Phys. }{\bf #1}(19#2)#3}
\def\prl#1#2#3{{\it Phys. Rev. Lett. }{\bf #1}(19#2)#3}

\def\pr#1#2#3{{\it Phys. Rev. }{\bf #1}(19#2)#3}
\def\np#1#2#3{{\it Nucl. Phys. }{\bf #1}(19#2)#3}

\relax

  \newcommand{\ccaption}[2]{
    \begin{center}
    \parbox{0.85\textwidth}{
      \caption[#1]{\small{\it{#2}}}
      }
    \end{center}
    }
\begin{document}
\newcommand\sss{\scriptscriptstyle}
\newcommand\mug{\mu_\gamma}
\newcommand\mue{\mu_e}
\newcommand\muf{\mu_{\sss F}}
\newcommand\mur{\mu_{\sss R}}
\newcommand\muo{\mu_0}
\newcommand\me{m_e}
\newcommand\as{\alpha_{\sss S}}
\newcommand\ep{\epsilon}
\newcommand\epb{\overline{\epsilon}}
\newcommand\aem{\alpha_{\rm em}}
\newcommand\refq[1]{$^{[#1]}$}
\newcommand\avr[1]{\left\langle #1 \right\rangle}
\newcommand\lambdamsb{\Lambda_5^{\rm \sss \overline{MS}}}
\newcommand\qqb{{q\overline{q}}}
\newcommand\qb{\overline{q}}
\newcommand\MSB{{\rm \overline{MS}}}
\newcommand\DIG{{\rm DIS}_\gamma}
\renewcommand\topfraction{1}       
\renewcommand\bottomfraction{1}    
\renewcommand\textfraction{0}      
\setcounter{topnumber}{5}          
\setcounter{bottomnumber}{5}       
\setcounter{totalnumber}{5}        
\setcounter{dbltopnumber}{2}       
\newsavebox\tmpfig
\newcommand\settmpfig[1]{\sbox{\tmpfig}{\mbox{\ref{#1}}}}
\begin{titlepage}
\nopagebreak
\vspace*{-1in}
{\leftskip 11cm
\normalsize
\noindent
\newline
CERN-TH/95-143 \newline
GEF-TH-5/1995 \newline
IFUM 506/FT \newline
hep-ph/yymmxxx

}
\vskip 1.0cm
\vfill
\begin{center}
{\large \bf \sc Differential Distributions for}

{\large \bf \sc Heavy Flavour Production at HERA}
\vfill
\vskip .6cm
{\bf Stefano Frixione}\footnote{Address after June 1: ETH, Zurich,
Switzerland}
\vskip .2cm
{Dip. di Fisica, Universit\`a di Genova, Genoa, Italy}\\
\vskip .5cm
{\bf Paolo Nason\footnotemark}
\footnotetext{On leave of absence from INFN, Sezione di Milano, Milan, Italy.}
\vskip .2cm
{CERN TH-Division, CH-1211 Geneva 23, Switzerland}
\vskip .5cm
{\bf Giovanni Ridolfi}
\vskip .2cm
{INFN, Sezione di Genova, Genoa, Italy.}

\end{center}
\vfill
\nopagebreak
\begin{abstract}
{\small
We compute pseudorapidity and transverse momentum distributions
for charm and bottom production at HERA.
We examine the effect of next-to-leading order QCD corrections, the
effect of possible intrinsic transverse momenta of the incoming
partons, and of fragmentation.
We compare our results with those of a full Monte Carlo simulation
using HERWIG.
The importance of the hadronic component of the photon is also studied.
We examine the possibility of
distinguishing between different parametrizations of the photon parton
densities
using charm production data, and the possibility of extracting
information about
the small-$x$ behaviour of the gluon density
of the proton. We also give a prediction for the transverse momentum and
pseudorapidity distributions for bottom production at HERA.
}
\end{abstract}
\vfill
CERN-TH/95-143/95 \newline
May 1995    \hfill
\end{titlepage}

\section{Introduction}
Experimental results on charm photoproduction at HERA have recently
become available [\ref{ZeusCharm}]. It is likely that more detailed
results on the differential distributions will appear in the
near future. With respect to previous photoproduction experiments
(ref.~[\ref{FixedTargetPhotoproduction}]) HERA offers the new opportunity
of a higher energy regime. A theoretical study of the total
photoproduction cross section has already been given in ref.~[\ref{FMNR_TOT}].
In the present work we extend the analysis of ref.~[\ref{FMNR_TOT}]
(to which we refer the reader for a general introduction and for notation) by
considering single inclusive distributions for
heavy flavour production at HERA.
Our analysis is based on the next-to-leading order calculation of
heavy-quark photoproduction and hadroproduction cross sections performed
in refs.~[\ref{EllisNason},\ref{NDE}], as implemented in a computer
program developed in refs.~[\ref{MNR},\ref{FMNRphoto}].
We will consider cross sections at fixed photon energies.
When necessary, a discussion of the energy dependence
of the distributions will be given.
We will also consider electroproduction for small photon
virtuality, in the  Weizs\"acker-Williams approximation, which is
appropriate for the bulk of the photoproduction cross section.
We will not consider the production of heavy flavours in deeply inelastic
events, i.e. events in which high photon virtuality is required
(see refs.~[\ref{VirtualPhoton}]).
The paper is organized as follows. In section~2 we study the point-like
contribution of the transverse
momentum and pseudorapidity distributions in photoproduction.
Section~3 is devoted to the study of the hadronic component,
considering the possibility of separating it from the point-like
component with appropriate cuts.
In sections~4 and 5 we consider electroproduction of charm and bottom,
in the Weizs\"acker-Williams approximation.
In section~6 we compare our fixed-order
results with those of the Monte Carlo HERWIG [\ref{HERWIG}], and in
section~7 we give our conclusions.
Some technical details on the factorization schemes, the scale dependence,
and the modified Weizs\"acker-Williams approximation used here are given
in the Appendix.

\section{point-like component}
\begin{figure}[tbhp]
  \begin{center}
    \mbox{
      \epsfig{file=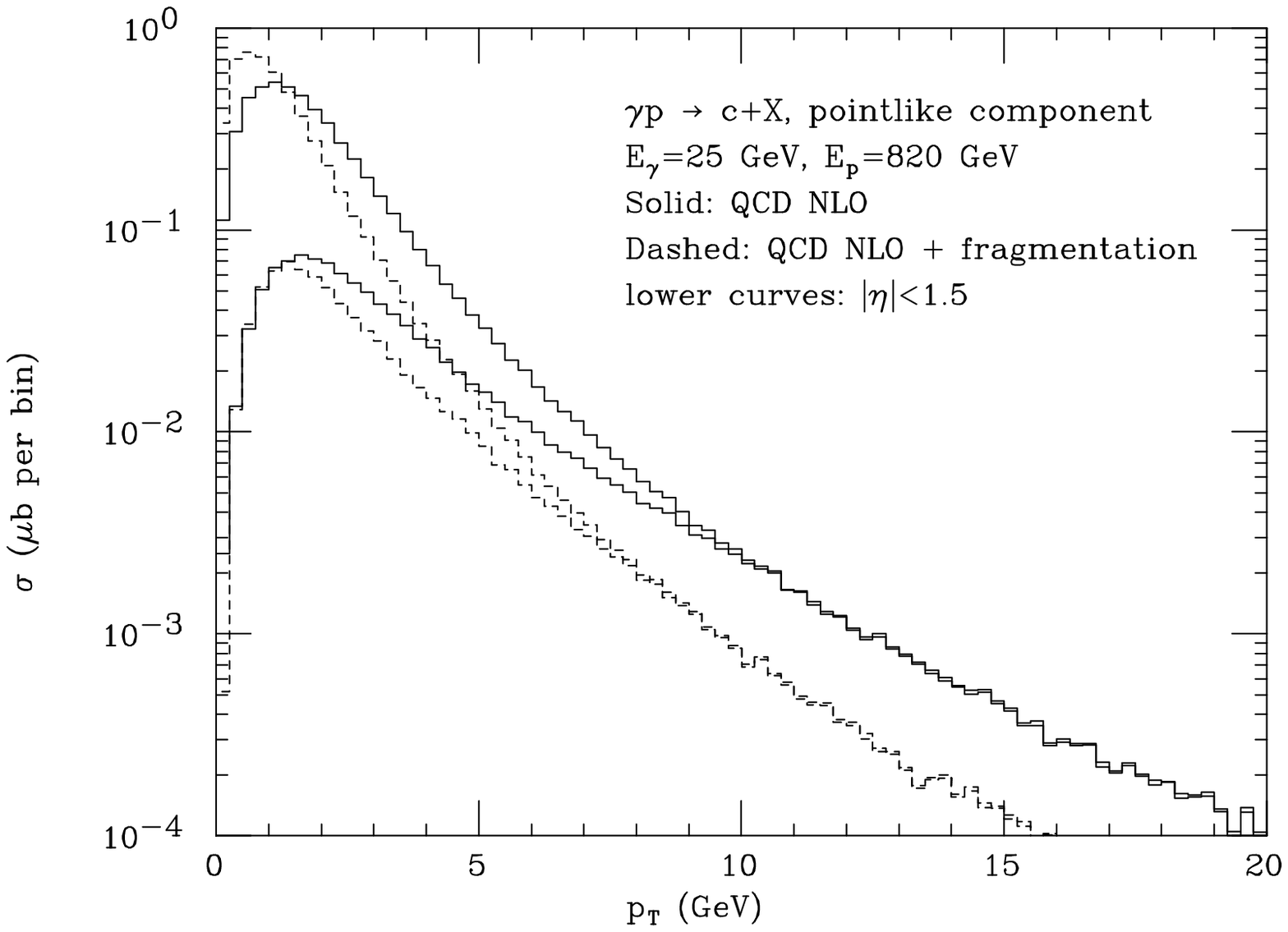,width=0.60\textwidth}
      }
  \ccaption{}{\label{pt286pnt}
 Charm transverse momentum distribution in photon-proton collisions,
 with and without a pseudorapidity cut.
 The effect of applying a Peterson fragmentation function
 to the final-state quark is also shown.
}
  \end{center}
\end{figure}

\begin{figure}[tbhp]
  \begin{center}
    \mbox{
      \epsfig{file=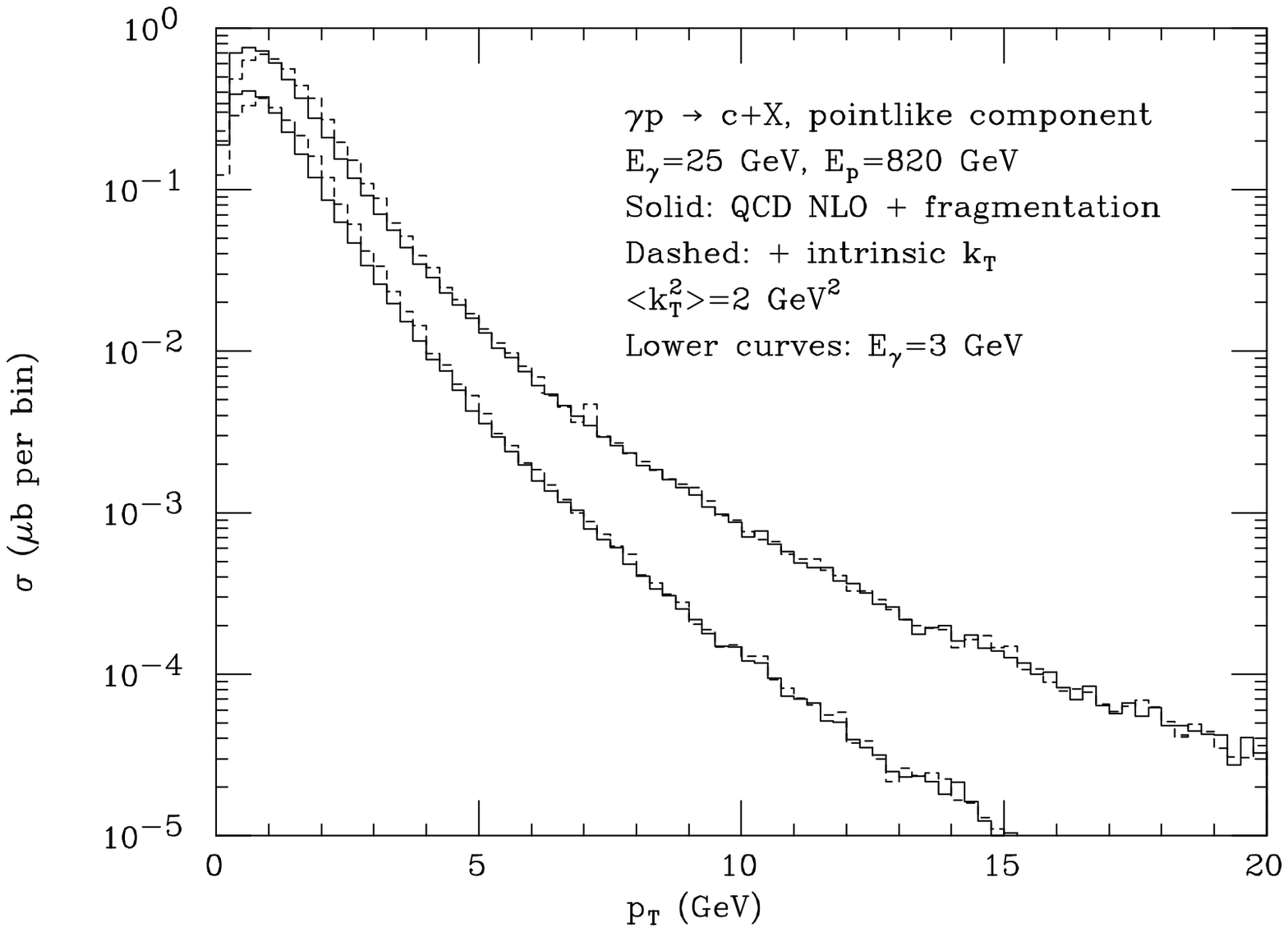,width=0.60\textwidth}
      }
  \ccaption{}{\label{ptkickpnt}
 Charm transverse momentum distribution in photon-proton collisions,
 for two values of the photon energy.
 The effect of applying an intrinsic transverse momentum to the incoming
 parton is also shown.
}
  \end{center}
\end{figure}

Unless specifically stated, we will use in the
following the set of parton densities MRSA, ref.~[\ref{MRSA}],
with $\Lambda_5=151\,$MeV. This set of distribution functions has
been recently updated (MRSG, ref.~[\ref{MRSG}]), to
include new HERA deep inelastic scattering data, allowing
a different small-$x$ behaviour of the gluon and sea quark densities.
We have checked that the shapes of the single-inclusive distributions
we are considering are not significantly different when the new parametrization
is used.
The default value of the charm quark mass
will be $m_c=1.5\,$GeV. The renormalization
scale will be taken as $\mur=\mu_0$, and the factorization scale for the proton
and for the photon will be taken as $\muf=\mug=2\mu_0$,
where $\mu_0=\sqrt{p_{\sss T}^2+m_c^2}$.
We begin by showing the transverse momentum
distribution for the charm quark in fig.~\ref{pt286pnt}, for $E_\gamma=25$~GeV
(the proton energy $E_p$ will be fixed in the following to be 820 GeV).
We also show the effect of applying the Peterson fragmentation
function~[\ref{pete}]
\beq
D(x)\,=\,\frac{1}{x\left(1-1/x-\epsilon/(1-x)\right)^2}
\eeq
to the final state quark. We use the value $\epsilon=0.06$, which is
the central value quoted in ref.~[\ref{Chrin}] for charm quarks.
As can be expected this softens considerably the $p_{\sss T}$ spectrum.
As discussed in ref.~[\ref{FMNRft}], in the case of
fixed target photoproduction experiments, the inclusion of Peterson
fragmentation correctly reproduces the shape of the measured $p_{\sss T}$
distribution.
The effect of a pseudorapidity cut similar to the one applied by the
ZEUS collaboration [\ref{ZeusCharm}] is also shown. At low to moderate
$p_{\sss T}$ this cut considerably lowers the cross section,
while at higher $p_{\sss T}$
it has a negligible effect. In fig.~\ref{ptkickpnt} we compare
the $p_{\sss T}$ distributions at different photon energies. We also illustrate
the effect of applying an intrinsic transverse momentum $k_T$
to the incoming parton (see ref.~[\ref{FMNRft}]).
We find that even with the very large value of
$\langle k_T^2 \rangle=2\,$GeV$^2$ the effect is small.
Therefore, in the following, we will
neglect the effect of an intrinsic transverse momentum of the incoming
partons.

\begin{figure}[tbhp]
  \begin{center}
    \mbox{
      \epsfig{file=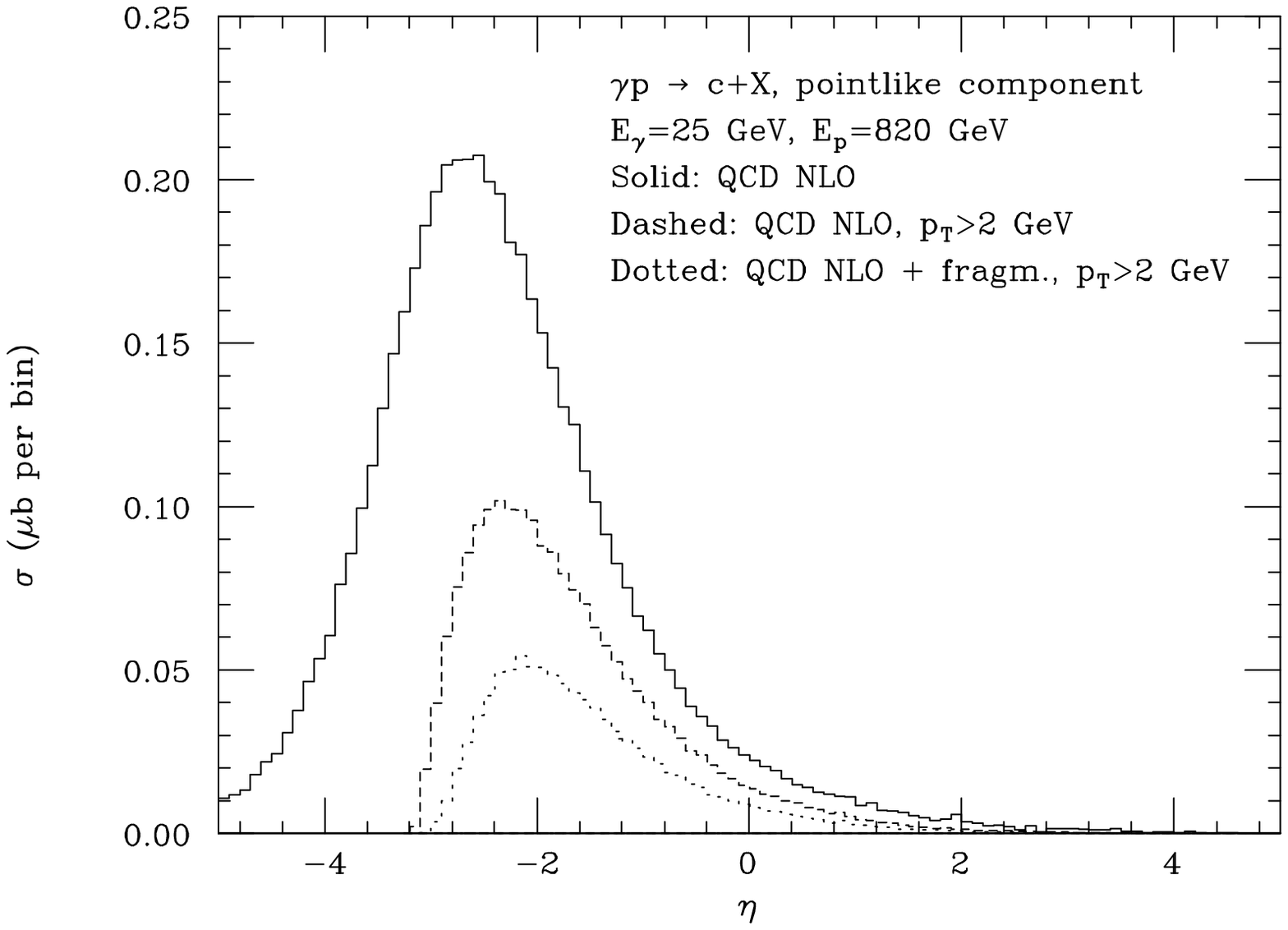,width=0.60\textwidth}
      }
  \ccaption{}{\label{eta286pnt}
 Charm pseudorapidity distribution in photon-proton collisions, with
 and without a transverse momentum cut. The effect of fragmentation
 is also shown.
}
  \end{center}
\end{figure}

\begin{figure}[tbhp]
\settmpfig{eta286pnt}
  \begin{center}
    \mbox{
      \epsfig{file=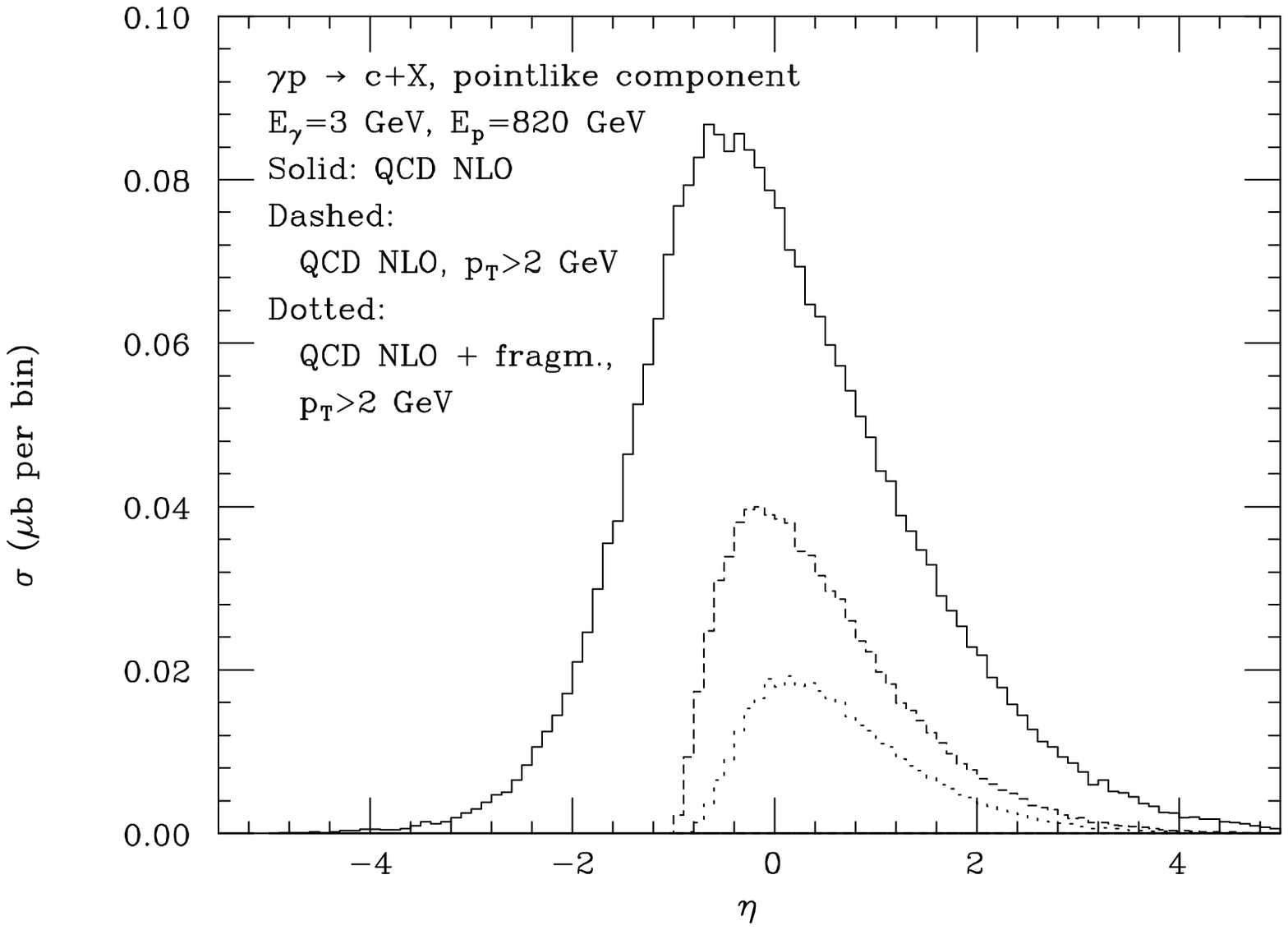,width=0.60\textwidth}
      }
  \ccaption{}{\label{eta99pnt}
 As in fig.~\box\tmpfig, for $E_\gamma=3\,$GeV.
}
  \end{center}
\end{figure}

Pseudorapidity distributions for charm are shown in fig.~\ref{eta286pnt}
for $E_\gamma=25$~GeV.
We observe that the point-like contribution to the cross section at this
energy strongly favours large negative pseudorapidities.
A cut on the transverse
momentum of the produced quark tends to move the distribution towards
the central region. The dotted curve shows the effect of fragmentation
in the presence of a transverse momentum cut. The fragmentation
has little effect on the pseudorapidity
of the quark, but degrades its transverse momentum, so that the
$p_{\sss T}$ cut has a stronger effect. The effect of fragmentation
without transverse momentum cut is not
shown in the figure, since it is not well defined (see ref.~[\ref{FMNRft}]).
At lower energies (fig.~\ref{eta99pnt}) the pseudorapidity distributions
become more central.

We now turn to the sensitivity of our distributions to
the various parameters that enter the computation. We have studied
the dependence of our distributions upon the charm quark mass,
by varying it between $1.2$ and $1.8$ GeV, on the renormalization
scale, which was varied by a factor of 2 below and above its
default value, and upon the distribution functions,
by considering the MRSD--$^\prime$ [\ref{MRSDmeno}] and the CTEQ2MF
[\ref{CTEQ2MF}] sets. The proton parton densities
were chosen in order to span the allowed range for the
small-$x$ behaviour of the gluon density, MRSD--$^\prime$
being the most singular, and CTEQ2MF the least singular one
(see ref.~[\ref{FMNR_TOT}] for a discussion of this point).
Studying the variation of our results with respect to $\Lambda$ is a difficult
problem,
since the values of $\Lambda$ extracted from fits to deep inelastic
scattering data are not in good agreement with the LEP value.
The CTEQ group has provided a parton density set in which the value
of $\Lambda$ was pushed as high as possible (see also ref.~[\ref{GRV_lam}]).
They use a value
of $\Lambda_5=220\,$MeV, which is still below the LEP value, but adequate
for a study of the sensitivity of the distributions. The set of parton
densities obtained in this fit (CTEQ2ML) was used in association with
this value of $\Lambda$.

\begin{figure}[tbhp]
  \begin{center}
    \mbox{
      \epsfig{file=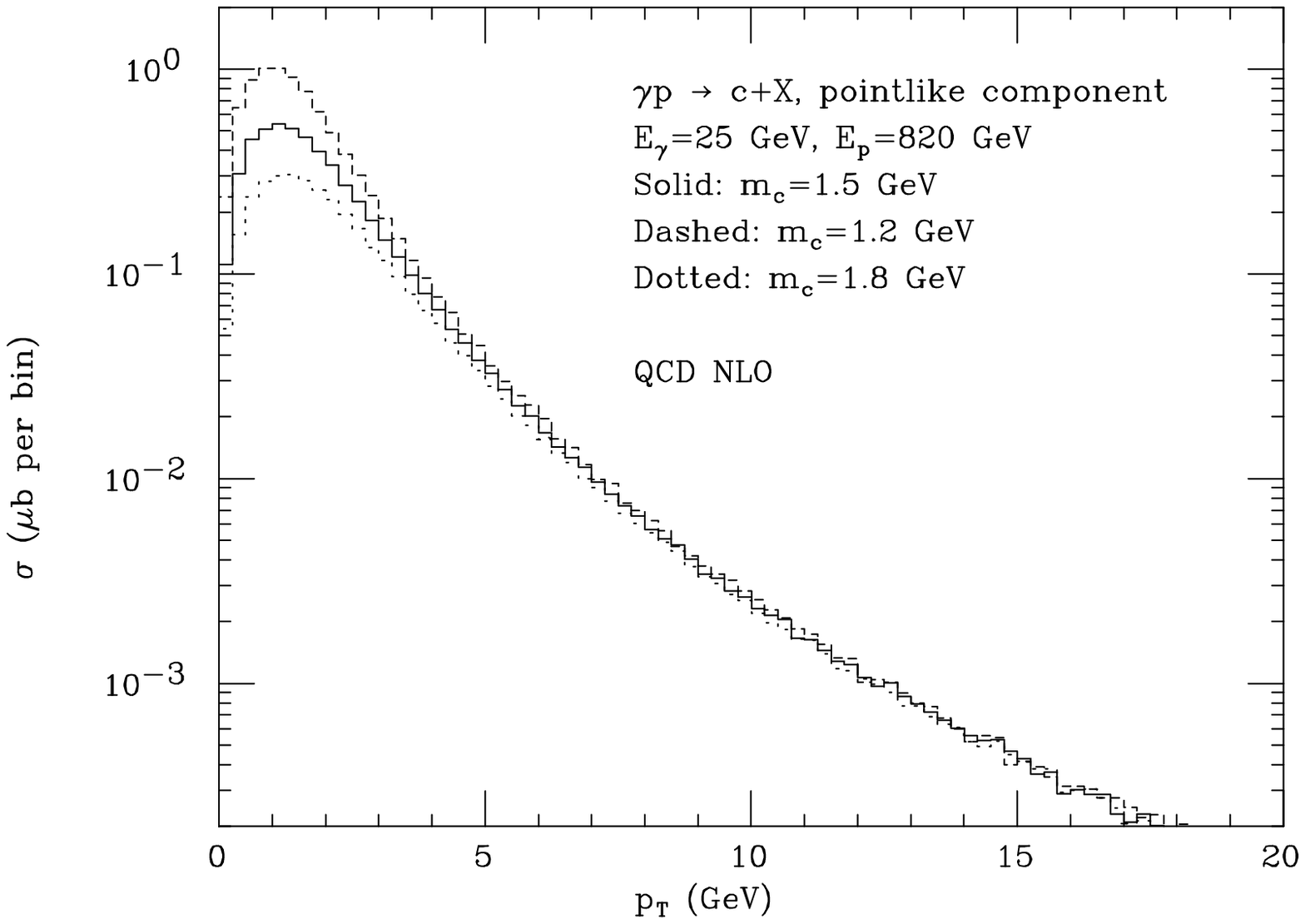,width=0.60\textwidth}
      }
  \ccaption{}{\label{pt286pnt_mvar}
 Mass dependence of the $p_{\sss T}$ distribution of charm.
}
  \end{center}
\end{figure}

\begin{figure}[tbhp]
  \begin{center}
    \mbox{
      \epsfig{file=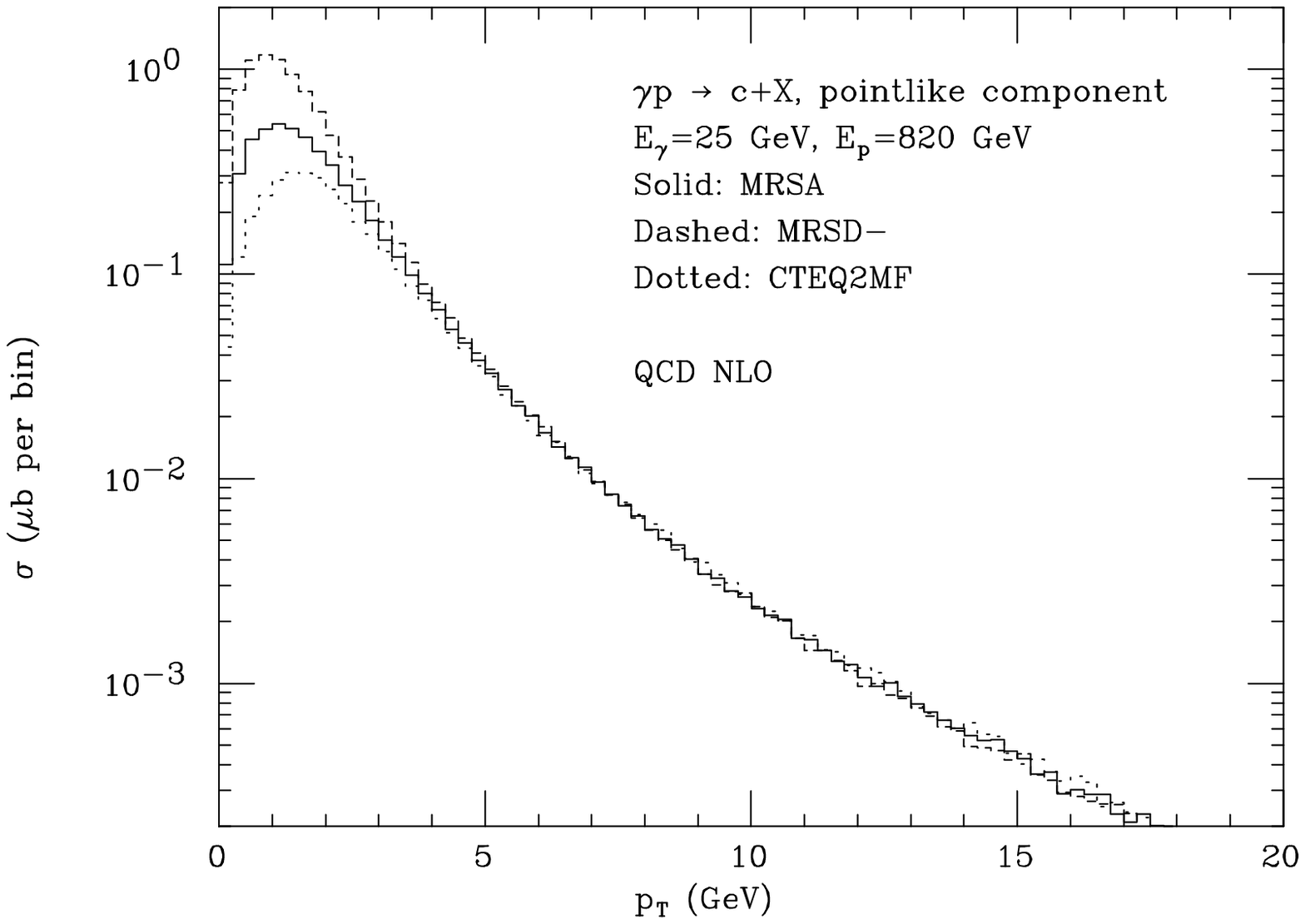,width=0.60\textwidth}
      }
  \ccaption{}{\label{pt286pnt_strfvar}
 Sensitivity of the $p_{\sss T}$ distribution of charm to the parton
 densities.
}
  \end{center}
\end{figure}

We found that the only cases in which there is
a noticeable shape variation is in the $p_{\sss T}$ distributions when we
vary the mass and the parton densities.
The relevant plots are given in figs.~\ref{pt286pnt_mvar} and
\ref{pt286pnt_strfvar}.
{}From the figures we can see a comparable variation of the $p_{\sss T}$
spectrum
in the low momentum region in the two cases. The mass dependence
is easily understood. We expect nearly no mass dependence at high
transverse momenta, while in the massless limit the cross section diverges
at small momenta. Therefore, the smaller the mass, the higher the cross
section at low $p_{\sss T}$. As far as the distribution function
dependence is concerned, the small-$x$ uncertainty in the densities
plays here a major r\^ole. The more singular the small-$x$ behaviour,
the higher the cross section at small $p_T$.

As a last point, we observe that, when measuring the total cross section,
if a transverse momentum cut on the $p_{\sss T}$ spectrum is being applied,
the extrapolated total cross section will depend upon the assumptions
we make about the value of the mass and the small-$x$ behaviour of
the distribution functions. This has to be properly taken into account
when inferring properties of the parton density
at small $x$ from the energy behaviour of the total charm cross section.

\section{Hadronic component}
\begin{figure}[tbhp]
  \begin{center}
    \mbox{
      \epsfig{file=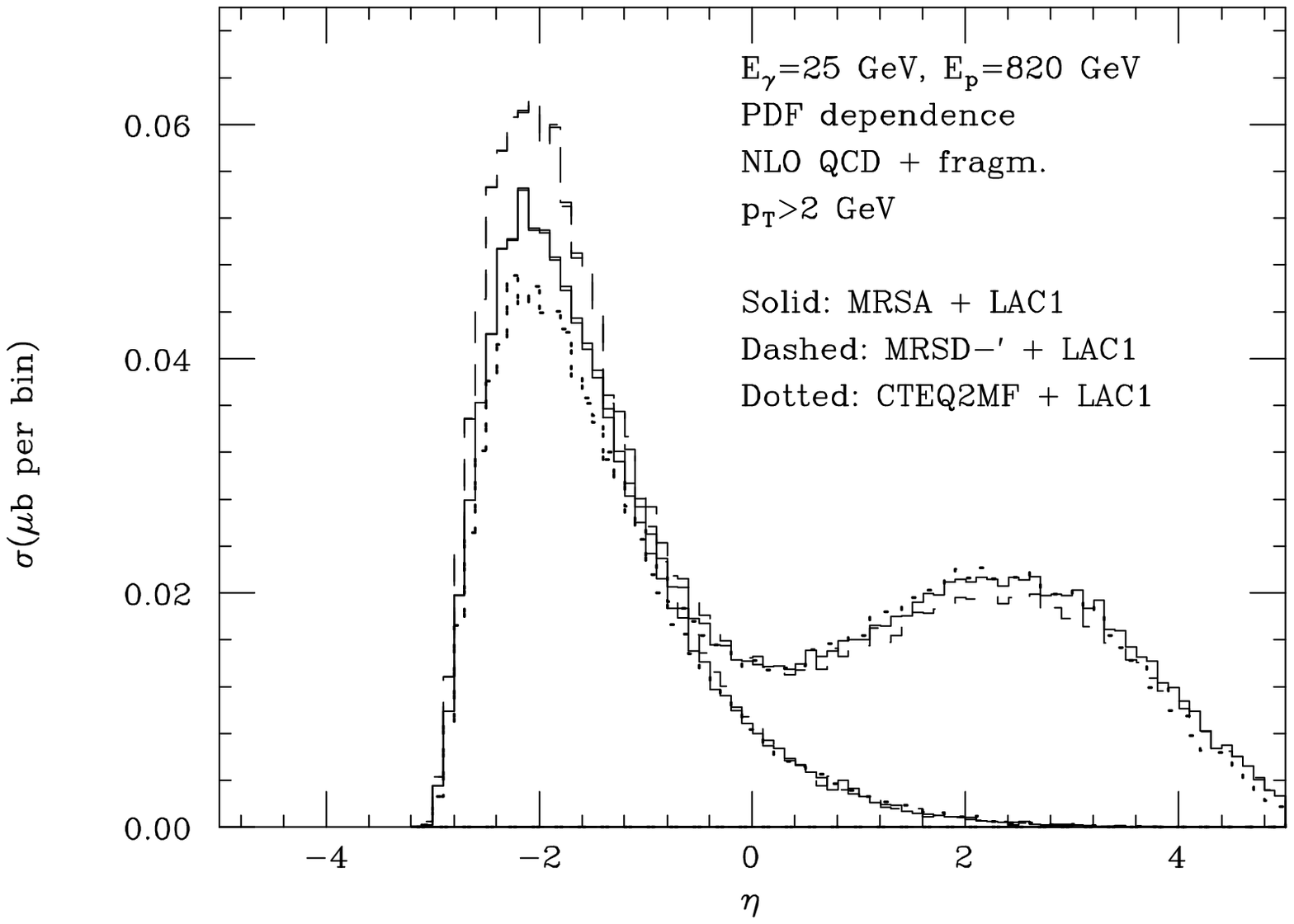,width=0.60\textwidth}
      }
  \ccaption{}{\label{pdf_LAC1_sum_286}
Pseudorapidity distribution of charm quarks,
obtained with the LAC1 set of photon parton densities, and three
different sets of proton parton densities. The point-like component is also
shown.
}
  \end{center}
\end{figure}

\begin{figure}[tbhp]
  \begin{center}
    \mbox{
      \epsfig{file=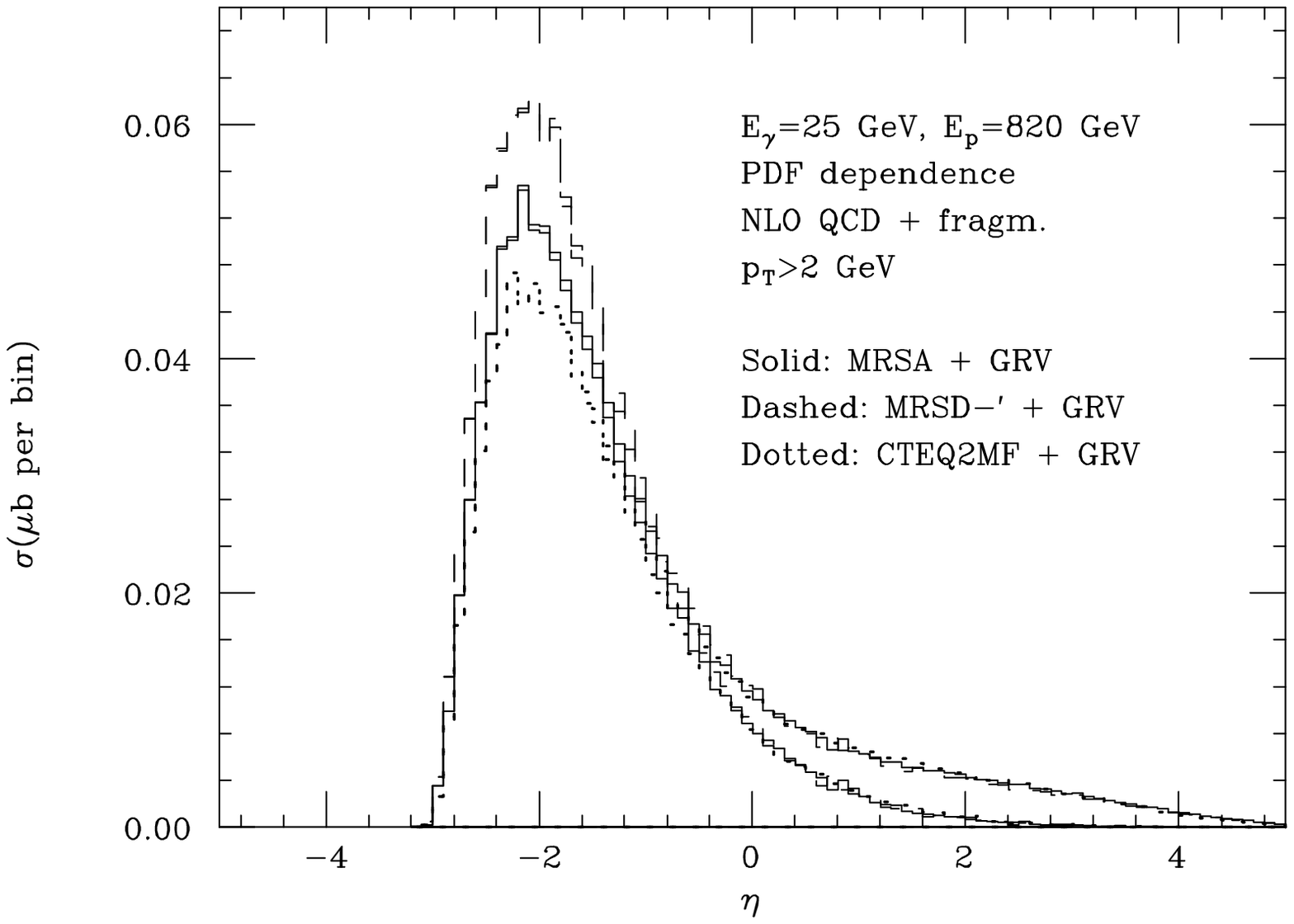,width=0.60\textwidth}
      }
  \ccaption{}{\label{pdf_GRV_sum_286}
Pseudorapidity distribution of charm quarks,
obtained with the GRV set of photon parton densities, and three
different sets of proton parton densities. The point-like component is also
shown.
}
  \end{center}
\end{figure}

The point-like contribution of the charm cross sections will be
contaminated by the hadronic component. Depending
upon the chosen distribution functions for the photon, the contribution
of the hadronic component to the total cross section may dominate
over the point-like one. Whatever the choice of the distribution functions,
however, the two components differ remarkably
in the pseudorapidity distribution. The point-like component,
as we have seen, favours negative pseudorapidities, while the hadronic
component favours the positive direction. The separation
of the two components will therefore require an analysis
of a large pseudorapidity range.
In the following, we will show a study of the full (i.e. point-like
plus hadronic) pseudorapidity distribution for different
choices of photon and proton distribution functions. We will see that
the shape and magnitude of the $\eta$ distribution will allow us to
distinguish between the most extreme parametrizations of photon
distribution functions.
In fig.~\ref{pdf_LAC1_sum_286} we show the $\eta$ distribution, obtained
with the LAC1~[\ref{LAC1}] set of photon parton densities, and three
different sets of proton parton densities; both the point-like and the
full results are presented.
A similar plot for the GRV~[\ref{GRVph}] set for the photon is given in
fig.~\ref{pdf_GRV_sum_286}.
We have chosen the LAC1 and GRV sets because, for our purposes, they represent
the two extreme possibilities, as discussed in ref.~[\ref{FMNR_TOT}].
In figs.~\ref{pdf_LAC1_sum_286} and \ref{pdf_GRV_sum_286} we have imposed
a realistic transverse momentum cut, which in general will make the
pseudorapidity distribution even narrower. In spite of the cut, the separation
of the two contributions is quite apparent.

The study of the pseudorapidity distribution could help in distinguishing
among different proton parton density sets, especially if the large negative
pseudorapidity region could be explored. On the other hand, the large
difference in shape induced by the two photon sets considered has
measurable effects even in the central region.

\begin{figure}[tbhp]
  \begin{center}
    \mbox{
      \epsfig{file=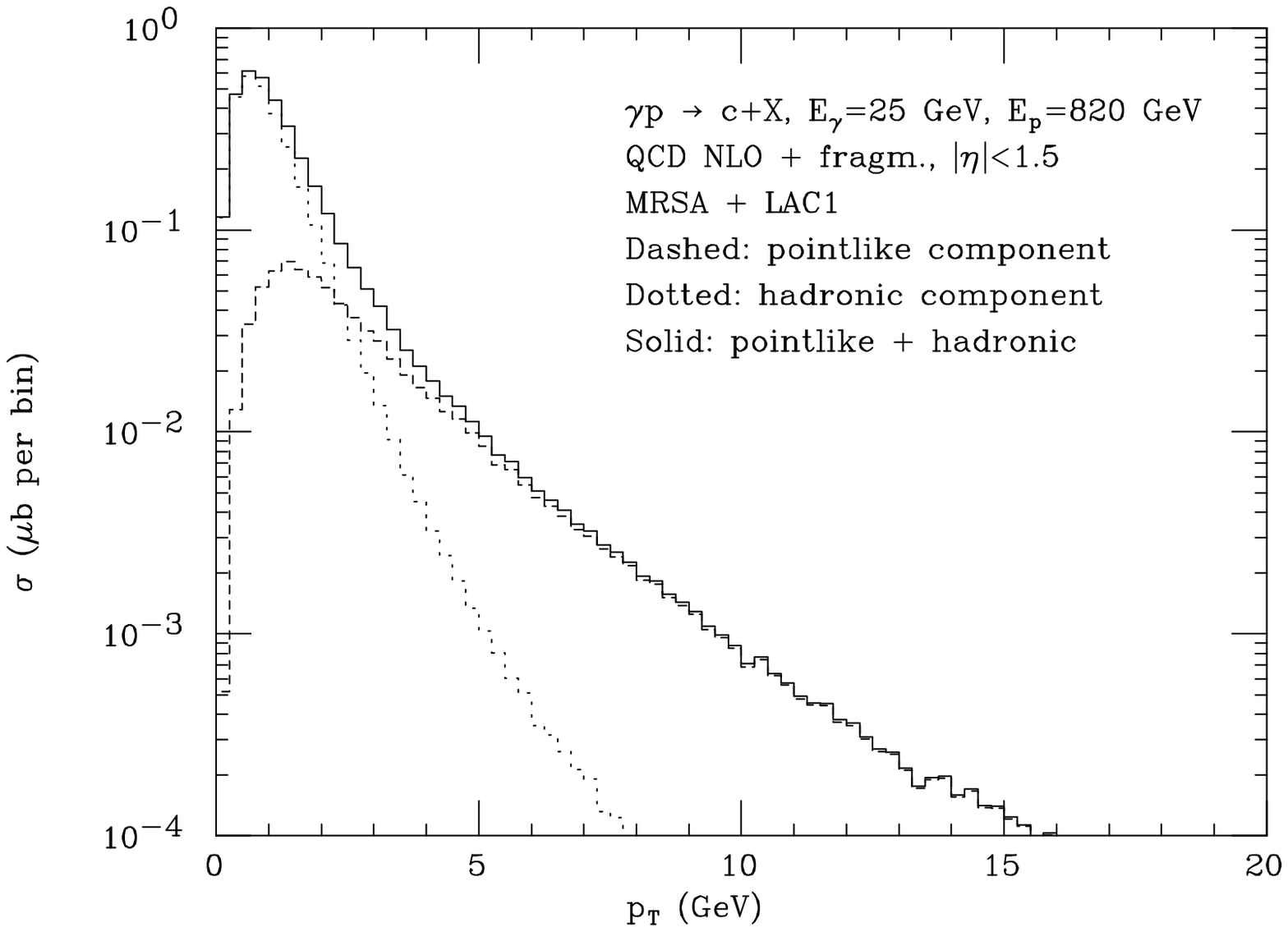,width=0.60\textwidth}
      }
  \ccaption{}{\label{pt286full}
Transverse momentum distribution of charm quarks.
The point-like and hadronic contributions are separately shown,
together with the sum of the two.
}
  \end{center}
\end{figure}

The full transverse momentum distribution is presented in
fig.~\ref{pt286full}. The point-like and hadronic contributions
are also separately shown. From the figure, we can see that
the hadronic contribution is dominant in the low-$p_{\sss T}$
region. At $p_{\sss T}\simeq 2$~GeV, it amounts to 50\%
of the full cross section. Therefore, even without investigating
the very low-$p_{\sss T}$ region, the softer behaviour of the full
cross section with respect to the point-like one should be visible.
Obviously, this behaviour is strongly influenced by
the photon parton densities used. If the GRV set were used,
instead of the LAC1 set, the full distribution would differ only
sligthly from the point-like one already at moderate $p_{\sss T}$.
We can conclude that the charm transverse momentum spectrum at large
values of the photon energy could help in further constraining
the parton densities in the photon. From fig.~\ref{pdf_LAC1_sum_286}, we
also learn that asymmetric pseudorapidity cuts may enhance or suppress
the effect of the hadronic component in the $p_{\sss T}$ spectrum, thus
providing another handle for its study.

\section{Charm electroproduction}
In this section we discuss single-inclusive distributions for charm production
in electron-proton collisions in the Weizs\"acker-Williams approximation. We
take the electron and proton energies to be $27.5$~GeV and $820$~GeV,
respectively.
Since the Weizs\"acker-Williams function peaks in the small-$x$ region,
the bulk of the contribution to the cross section will be due to photons
of relatively low energy. Therefore, the r\^ole of the hadronic component
is less important with respect to the monochromatic photon case.
By applying a small-$p_{\sss T}$ cut, the contribution of the
hadronic component is even more suppressed, especially for
soft photon parton densities. Therefore, the QCD predictions at
next-to-leading order are less affected by the uncertainty originating
from the photon distribution functions.

\begin{figure}[tbhp]
  \begin{center}
    \mbox{
      \epsfig{file=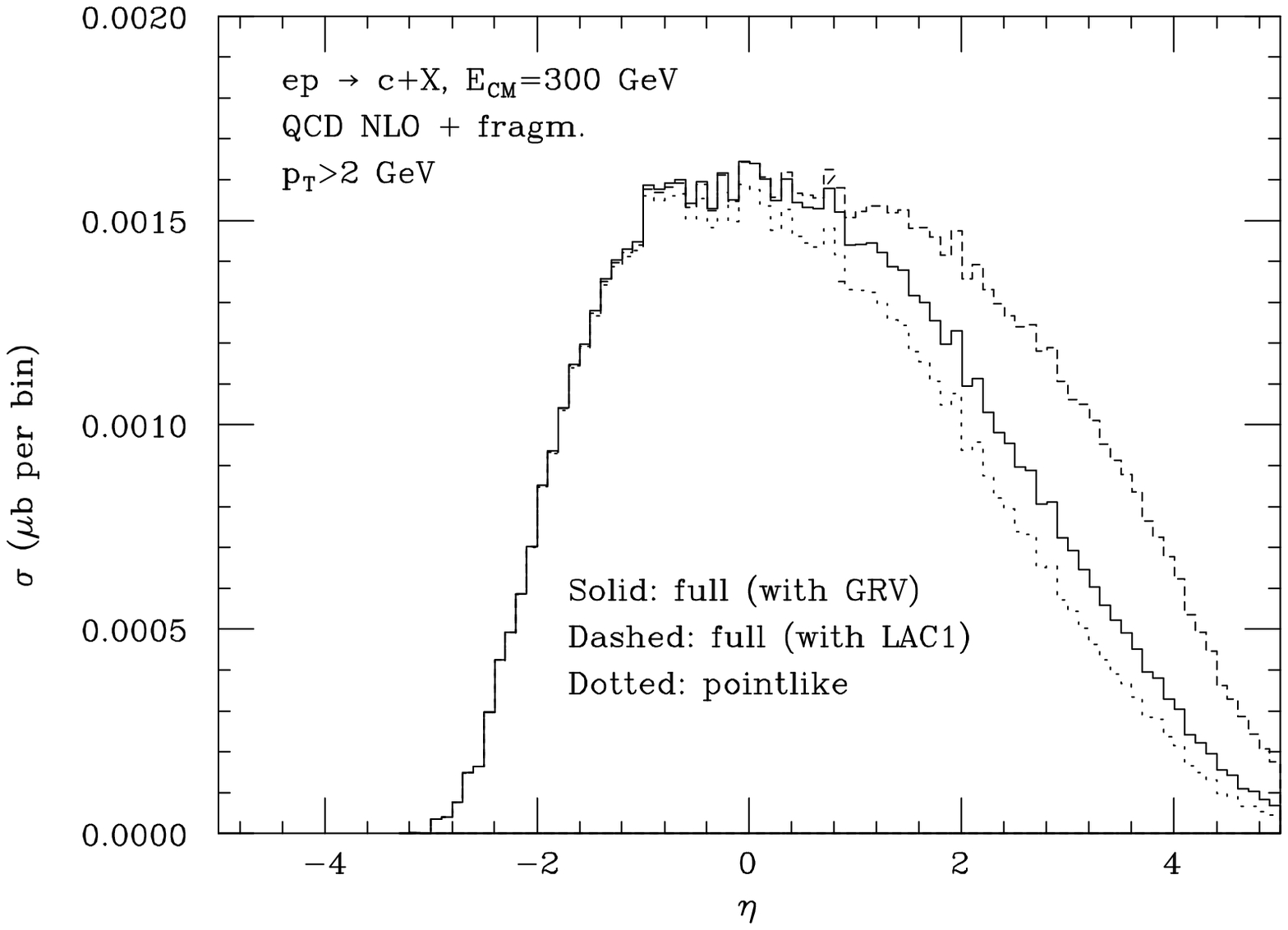,width=0.60\textwidth}
      }
  \ccaption{}{\label{el_eta_all}
Pseudorapidity distribution for charm electroproduction. The proton parton
density set MRSA is used.
}
  \end{center}
\end{figure}

\begin{figure}[tbhp]
  \begin{center}
    \mbox{
      \epsfig{file=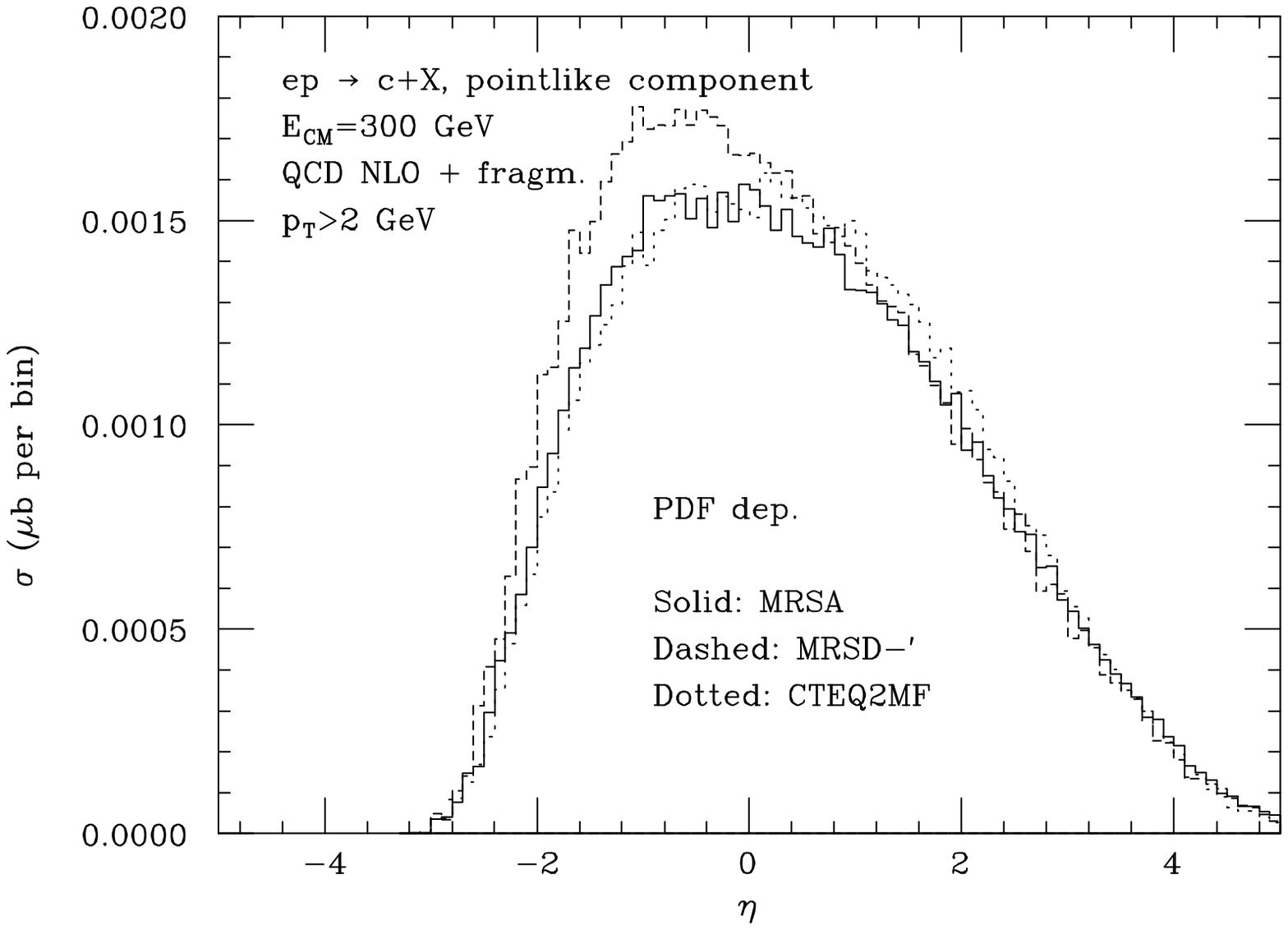,width=0.60\textwidth}
      }
  \ccaption{}{\label{el_eta_pdf}
Proton parton density dependence of the pseudorapidity distribution for
charm electroproduction; only the point-like component is shown.
}
  \end{center}
\end{figure}

Because of the softness of the incoming
photons, the pseudorapidity distribution of the point-like component is
concentrated in the central region. This can be seen in
fig.~\ref{el_eta_all}, where we show the pseudorapidity distribution for charm
quarks, supplemented with Peterson fragmentation and a transverse momentum cut.
The hadronic component is computed using our two extreme sets of photon parton
densities. It is apparent that its effect is small
in the central pseudorapidity region explored at HERA.
The curves in fig.~\ref{el_eta_all} are obtained with $m_c=1.5$~GeV,
$\mur=\mu_0$, $\muf=\mug=2\mu_0$ and the MRSA set for proton parton densities.
While the total cross section is quite sensitive to the value of these
parameters, as shown in ref.~[\ref{FMNR_TOT}], the shape of the
single-inclusive distributions is rather stable. The pseudorapidity
distribution shape is only mildly dependent upon the choice of proton parton
densities. This is due to the fact that the various parametrizations differ in
the small-$x$ region, which is not deeply probed in the electroproduction
process. This is shown, for the point-like component, in fig.~\ref{el_eta_pdf}.
The proton parton density dependence of the hadronic component is completely
negligible, due to the softness of the partons in the electron.

\begin{figure}[tbhp]
  \begin{center}
    \mbox{
      \epsfig{file=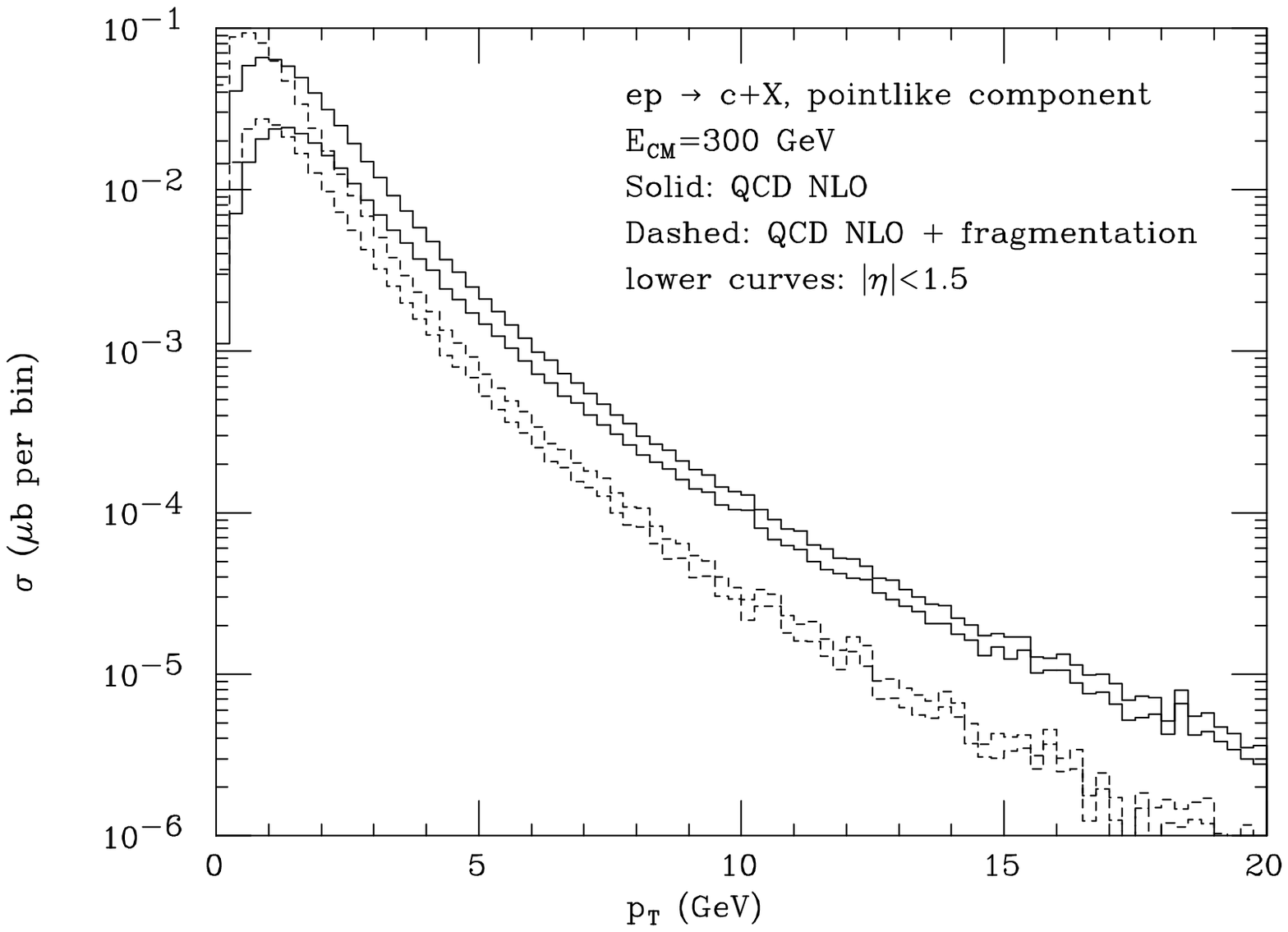,width=0.60\textwidth}
      }
  \ccaption{}{\label{el_pt_pnt}
Transverse momentum distribution for charm electroproduction (point-like
contribution).
}
  \end{center}
\end{figure}

We now consider the transverse momentum distribution.
In fig.~\ref{el_pt_pnt} we show the point-like contribution to
this differential cross section, with and without Peterson fragmentation. The
softening of the distribution due to fragmentation should be observable at
HERA.
The effect of applying a pseudorapidity cut is also shown: it
affects the shape of the distribution less dramatically than in the case
discussed in section 2.
We have checked that the hadronic component contribution to this
distribution is remarkably softer than the point-like one.
In practice, its effect is less than 10\% for $p_{\sss T} > 2$~GeV
when fragmentation is included and the pseudorapidity cut
$|\eta|<1.5$ is applied. We have also computed the same distribution for
different values of the renormalization and factorization scales, and we
found that the corresponding shape variations are small.

\begin{figure}[tbhp]
  \begin{center}
    \mbox{
      \epsfig{file=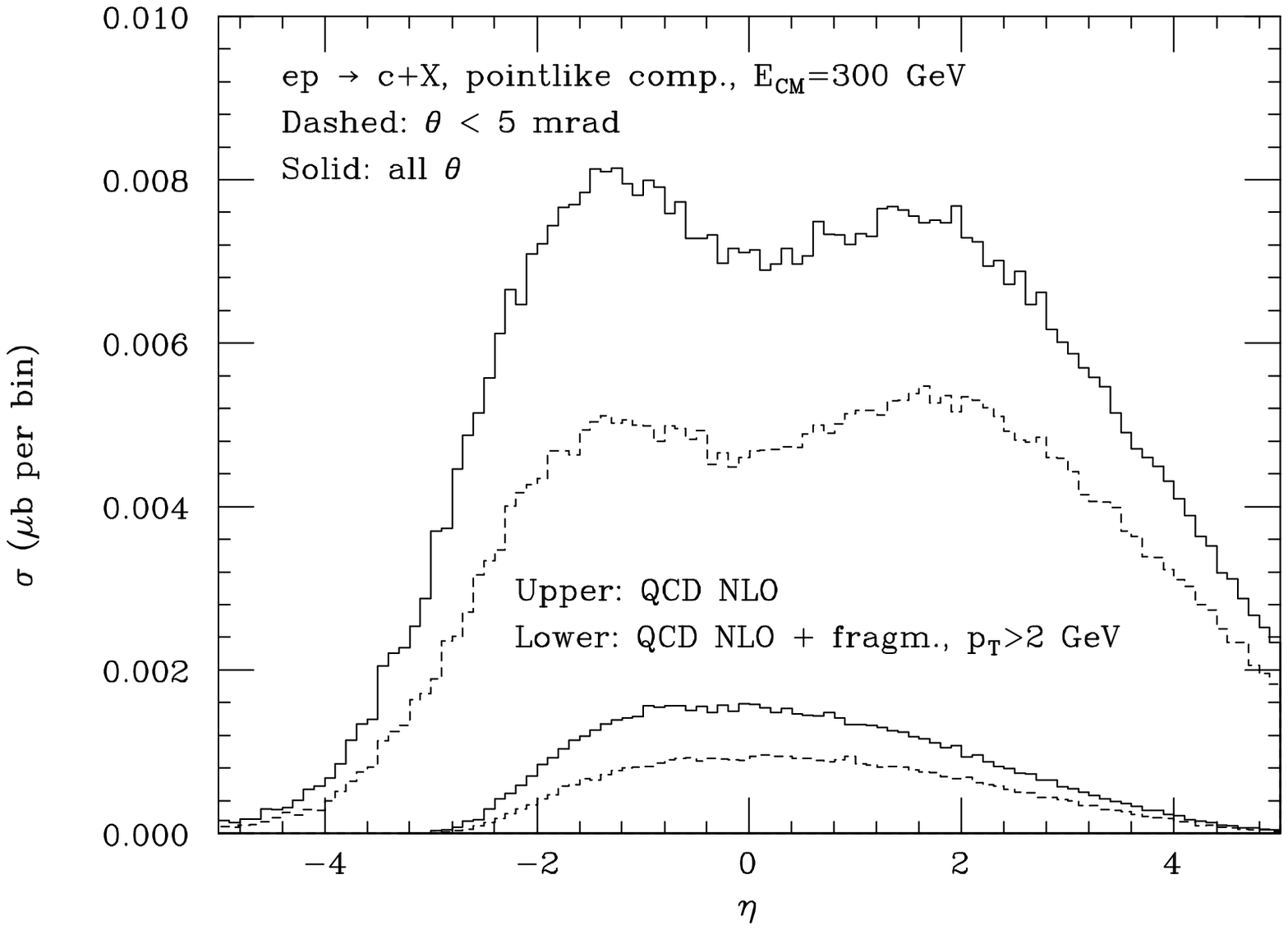,width=0.60\textwidth}
      }
  \ccaption{}{\label{el_eta_cut_pnt}
Pseudorapidity distribution for charm electroproduction, with and without
an antitag condition on the scattered electron.
}
  \end{center}
\end{figure}

All the distributions presented so far were also evaluated in the case when
an antitag condition on
the outgoing electron is applied, that is to say when the electron
scattering angle is below a given value, which we
chose to be 5~mrad. For this computation, the improved form of the
Weizs\"acker-Williams distribution presented in ref.~[\ref{FMNRww}] is
appropriate (see the appendix).
Differences in the shapes of the distributions are found to be
small. As an example, we show in fig.~\ref{el_eta_cut_pnt} the pseudorapidity
distribution with and without the antitag condition described above. The
antitag condition has the effect of decreasing the total cross section, as
described in ref.~[\ref{FMNR_TOT}], of slightly enhancing the contribution
of positive pseudorapidities, and of mildly softening the transverse
momentum spectrum.

We can conclude that charm electroproduction distributions are rather
insensitive to the choice of the parameters entering the perturbative
calculation.
Therefore, single-inclusive charm electroproduction will be of little help in
constraining the parton distribution functions of both the proton and
the photon.
On the other hand, the comparison between data and theoretical predictions
will be useful for the study of the production mechanism.

\section{Bottom electroproduction}
\begin{figure}[tbhp]
  \begin{center}
    \mbox{
      \epsfig{file=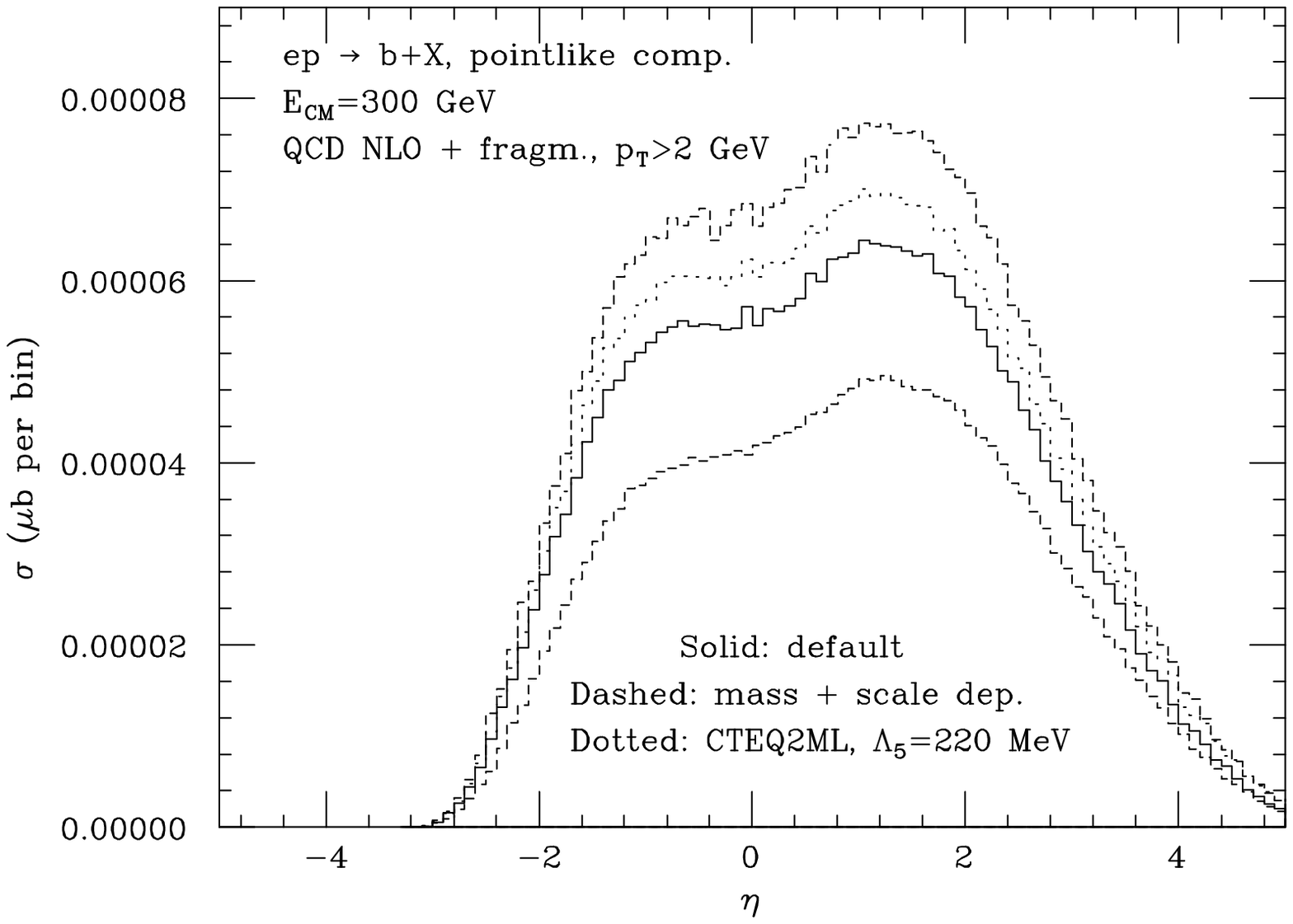,width=0.60\textwidth}
      }
  \ccaption{}{\label{b_el_eta_var}
Pseudorapidity distribution for bottom electroproduction
(point-like component only), with Peterson fragmentation
and a transverse momentum cut.
}
  \end{center}
\end{figure}

Due to the higher value of the quark mass, perturbative QCD predictions
for bottom production are more reliable than those for charm.
Furthermore, we are able to study the scale dependence
in an exhaustive manner. In fact, the
factorization scale, when varied by a factor of two below and
above its default value, is always higher than the mininum allowed
by the distribution function parametrizations.
In fig.~\ref{b_el_eta_var} we show the point-like component of
the pseudorapidity distribution for bottom electroproduction.
The solid curve represents our central prediction, which corresponds
to choosing $m_b=4.75$~GeV, $\mur=\muf=\mug=\mu_0$, where
$\mu_0=\sqrt{p_{\sss T}^2+m_b^2}$.
The QCD result is supplemented with Peterson fragmentation with
$\epsilon=0.006$, the central value of ref.~[\ref{Chrin}] for bottom
quarks.
In the case of bottom the effect of fragmentation is less important than in the
case of charm.
The band between the two dashed curves
has been obtained by varying the bottom quark mass between 4.5 and 5~GeV
and the renormalization and factorization scales between $\mu_0/2$ and
$2\mu_0$.
The proton parton density set chosen is MRSA.
We also show (dotted curve) the result obtained with a parton density set
fitted at a higher value
of $\Lambda_{\sss QCD}$, namely the set CTEQ2ML with $\Lambda_5=220$~MeV,
and the central values of mass and scales.
It is difficult to foresee what a realistic transverse
momentum cut could be for bottom production, and we have therefore kept the
same
cut $p_{\sss T}>2$~GeV we used in the charm case.

Figure~\ref{b_el_eta_var} does not describe the uncertainties
on the theoretical prediction completely, because the CTEQ2ML set
was used keeping
the scales and the mass at their default value, and because the hadronic
contribution was not included.
\begin{figure}[tbhp]
  \begin{center}
    \mbox{
      \epsfig{file=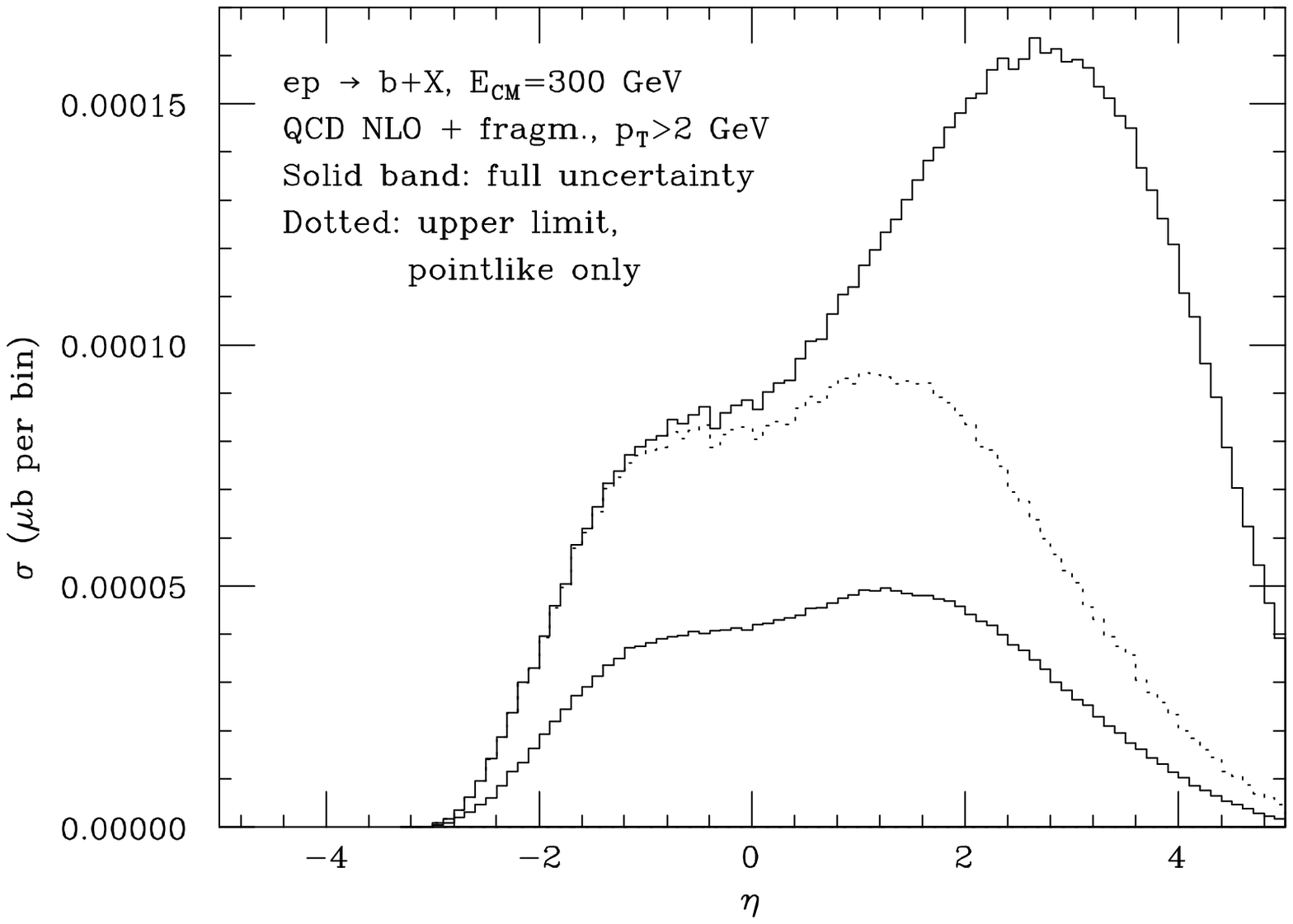,width=0.60\textwidth}
      }
  \ccaption{}{\label{b_el_eta_all}
Full uncertainty on the pseudorapidity distribution for bottom
electroproduction with Peterson fragmentation and a transverse momentum
cut.
}
  \end{center}
\end{figure}
The full uncertainty on the $\eta$ distribution is presented in
fig.~\ref{b_el_eta_all}, in which mass and scales are varied
also when using the CTEQ2ML set, and the possibility is considered
that the effect of the hadronic component is as large as the LAC1 set
implies, or is completely absent.
The dramatic effect of the inclusion of the hadronic component
is apparent in the positive pseudorapidity region. This is due
to the very soft behaviour of the LAC1 partons and to the stronger
sensitivity of the hadronic component to the choice of mass and
scales with respect to the one of the point-like component.

\begin{figure}[tbhp]
  \begin{center}
    \mbox{
      \epsfig{file=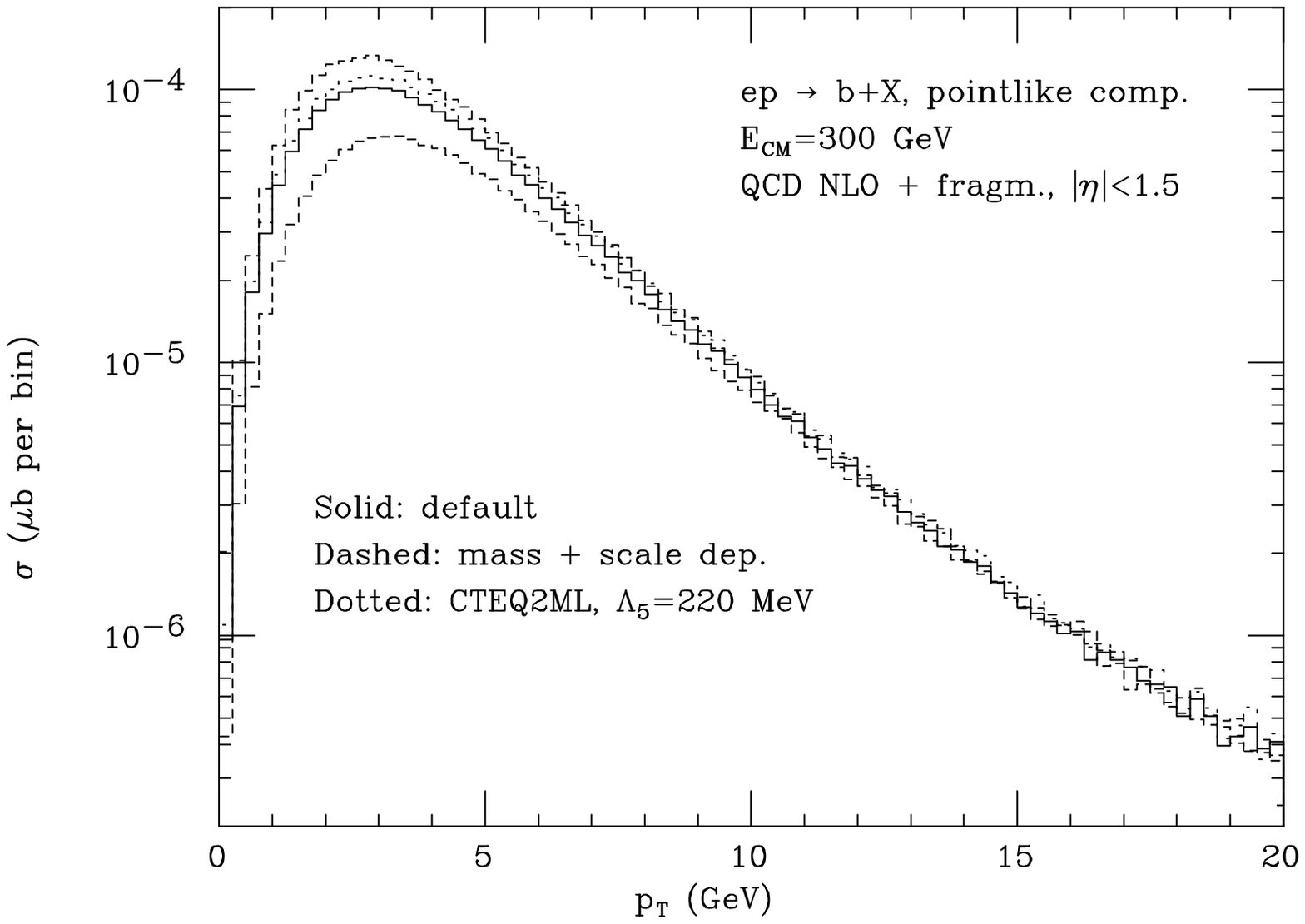,width=0.60\textwidth}
      }
  \ccaption{}{\label{b_el_pt_var}
Transverse momentum distribution for bottom electroproduction
(point-like component only) with Peterson fragmentation and a pseudorapidity
cut.
}
  \end{center}
\end{figure}

We now turn to the transverse momentum distribution, shown in
fig.~\ref{b_el_pt_var}, which is analogous to the corresponding
figure for the $\eta$ distribution, fig.~\ref{b_el_eta_var}.
As expected, in the large-$p_{\sss T}$ region the sensitivity to mass
and scales
choice is strongly reduced. Therefore, the perturbative prediction becomes
more reliable there.

\begin{figure}[tbhp]
  \begin{center}
    \mbox{
      \epsfig{file=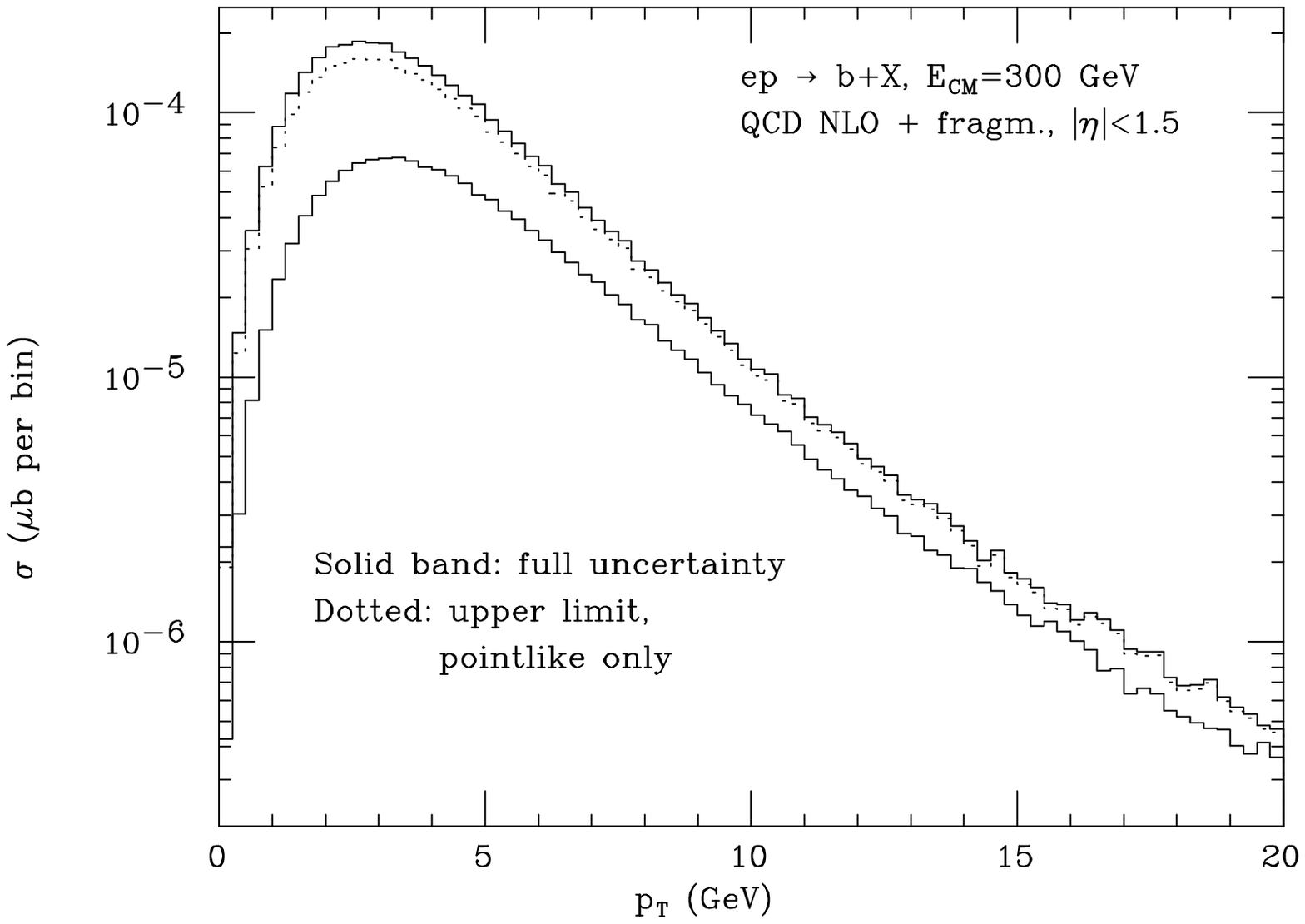,width=0.60\textwidth}
      }
  \ccaption{}{\label{b_el_pt_all}
Full uncertainty on the transverse momentum distribution for bottom
electroproduction with Peterson fragmentation and a transverse momentum
cut.
}
  \end{center}
\end{figure}

The full uncertainty on the $p_{\sss T}$ distribution is given in
fig.~\ref{b_el_pt_all}, which is analogous to fig.~\ref{b_el_eta_all}.
{}From the figure, it is quite clear that
even with the LAC1 set the hadronic component affects the prediction
only marginally. This fact is a consequence of the applied pseudorapidity cut,
as can be inferred from fig.~\ref{b_el_eta_all}.
We can therefore regard fig.~\ref{b_el_pt_all} as a reliable prediction
of QCD for the $p_{\sss T}$ spectrum of $b$ hadrons at HERA.

\section{Comparison with Monte Carlo results}
In order to assess the influence of higher-order and non-perturbative
contributions on the shape of single-inclusive distributions,
we have compared our fixed-order calculations with the results
obtained with the parton-shower Monte Carlo program HERWIG~[\ref{HERWIG}].
When using HERWIG we have always excluded the flavour excitation
processes. This is because comparable processes, such as gluon
splitting, are difficult to include. We have therefore preferred
to use the Monte Carlo at a consistent level of accuracy.
As discussed in section~2, the effect of an intrinsic transverse momentum
of the incoming partons is negligible in photoproduction at HERA energies.
This leads us to expect that higher-order and non-perturbative corrections
are not as important here as in the case of hadroproduction~[\ref{FMNRft}].
On the other hand, it is interesting to investigate how Peterson
fragmentation compares with the model of cluster hadronization
implemented in HERWIG. We limit our discussion to the point-like
component of the photoproduction cross section for fixed photon energy.

\begin{figure}[tbhp]
  \begin{center}
    \mbox{
      \epsfig{file=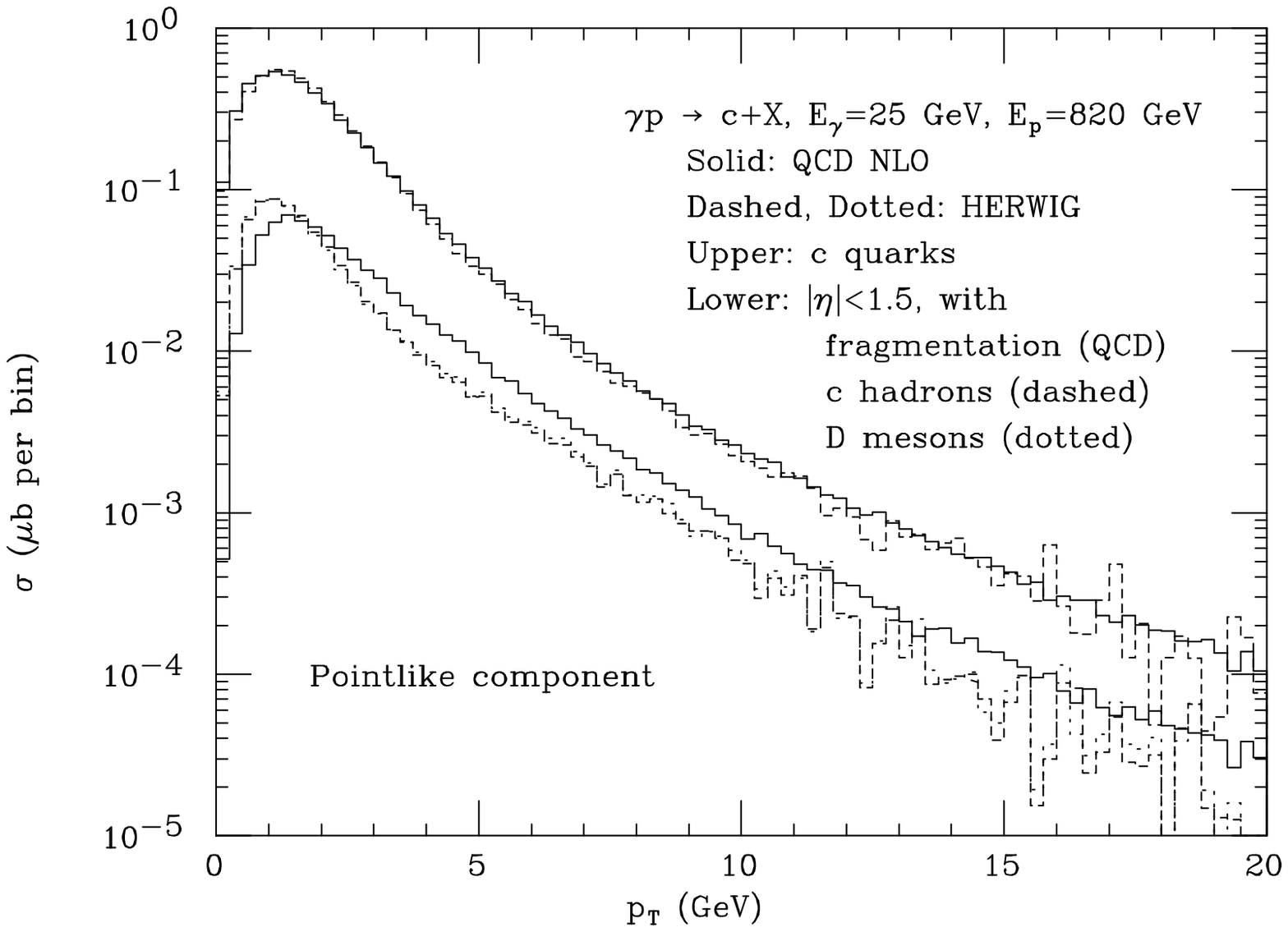,width=0.60\textwidth}
      }
  \ccaption{}{\label{pt_vs_hrwg}
Transverse momentum distribution for charm photoproduction
(point-like component) as given by next-to-leading order QCD and by the
Monte Carlo HERWIG.
}
  \end{center}
\end{figure}

In fig.~\ref{pt_vs_hrwg} we show the transverse momentum distribution
computed in fixed-order perturbation theory and with HERWIG. All HERWIG
curves have been normalized to give the same total cross section of the
corresponding next-to-leading order QCD predictions.
We observe that before the hadronization process is switched on, the QCD
prediction and the HERWIG result agree almost perfectly in all the presented
$p_{\sss T}$ range (upper curves). We then consider the effect of hadronization
in the central pseudorapidity region ($|\eta|<1.5$) by comparing the
perturbative prediction supplemented with Peterson fragmentation and the
HERWIG results for charmed hadrons or $D$ mesons only (lower curves).
In this case the two predictions display a different behaviour only in the
low-$p_{\sss T}$ region. This is also the only part of the $p_{\sss T}$
spectrum that is significantly affected by the pseudorapidity cut.
The results in fig.~\ref{pt_vs_hrwg} allow us to conclude that
Peterson fragmentation gives a description of the hadronization process,
which is consistent with the one suggested by cluster models only
at sufficiently large transverse momenta.

\begin{figure}[tbhp]
  \begin{center}
    \mbox{
      \epsfig{file=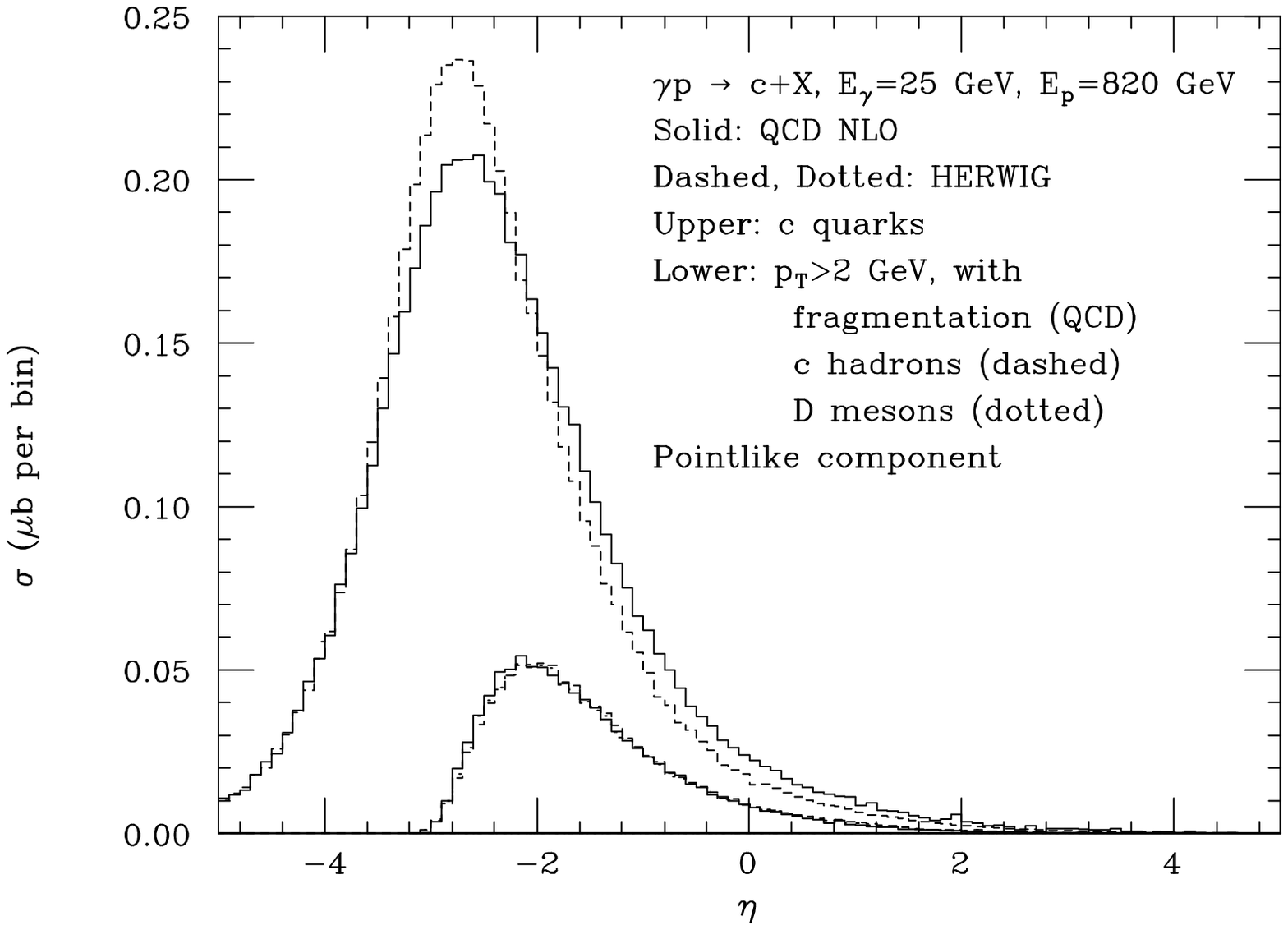,width=0.60\textwidth}
      }
  \ccaption{}{\label{eta_vs_hrwg}
Pseudorapidity distribution for charm photoproduction
(point-like component) as given by next-to-leading order QCD and by the
Monte Carlo HERWIG.
}
  \end{center}
\end{figure}

A similar study for the pseudorapidity distribution is presented in
fig.~\ref{eta_vs_hrwg}.
In this case, some difference is observed between the two distributions
before hadronization, the fixed-order QCD prediction being slightly
broader. However, they are peaked around the same value of $\eta$, which
is rather far from the central region, where the experimental acceptance is
larger.
Applying a small transverse momentum cut, the use of Peterson fragmentation
becomes possible also for the pseudorapidity distribution. We are therefore
able to compare the QCD next-to-leading order distribution, convoluted with
the Peterson fragmentation function, and the HERWIG curves for charmed hadrons.
We find a complete agreement between the two.

We have performed the same kind of analysis on the hadronic component, and
we found that the qualitative behaviour is completely analogous
to what has been presented for the point-like component.

\section{Conclusions}
We have studied single-inclusive cross sections for charm and
bottom production at HERA. We considered charm
production in monochromatic photon-proton collisions,
for energies spanning the range accessible at HERA.
We found that the shape
of the $p_{\sss T}$ and pseudorapidity distributions are quite stable
with respect to the variation of the input parameters
entering the calculation. On the other hand, the total
cross section value (see ref.~[\ref{FMNR_TOT}] for
a discussion on this point) is strongly affected by the choice
of parameters.
We found that from the study of the pseudorapidity distributions
in the central $\eta$ region ($|\eta|<1.5$) it is difficult to distinguish
between the various proton parton densities with different small-$x$
behaviour.
On the other hand, the striking
difference between the predictions obtained by using the
LAC1 and GRV set for photon distribution functions is clearly
visible.
This fact is almost completely due to the very soft behaviour
of the LAC1 parton densities.

The transverse momentum distribution for large $p_{\sss T}$
is reliably predicted by QCD. Measurements in this region
should constitute a good test of the production mechanism.
At moderate $p_{\sss T}$, the point-like result is strongly
influenced by the hadronic contribution when the LAC1 set for photon
parton densities is used. Experimental investigations in this region of
the spectrum could therefore help to distinguish
among the various parton distribution functions in the photon.

Charm electroproduction is a less clean test of the production
mechanism with respect to monochromatic photon production, because of
the convolution with the Weizs\"acker-Williams function.
On the other hand, the softness of the latter implies a strongly
reduced dependence of the total cross section upon the phenomenological
parameters entering the calculation. Furthermore, when a small
transverse momentum cut is applied, the hadronic contribution
is more suppressed than the point-like one, whatever the photon
densitites used. Therefore, electroproduction is less sensitive
to the contamination of the hadronic component than photoproduction
with a monochromatic photon, and can be used to further investigate
the point-like production mechanism. Unfortunately, a sizeable dependence
upon charm mass and renormalization scale remains in the total cross section
prediction.

The case in which an antitag condition is applied on the outgoing
electron was also studied. Aside from the effect on the total
cross section value (which has already been discussed in
ref.~[\ref{FMNR_TOT}]), the angular cut slightly enhances the importance
of the central pseudorapidity region and mildly softens the
transverse momentum spectrum.

We also considered the production of bottom quarks in $ep$ collisions.
Due to the higher value of the quark mass, perturbative QCD predictions
for bottom production are more reliable than those for charm.
It turns out that
the point-like component is quite stable in shape.
The hadronic component has a dramatic effect on the pseudorapidity
distribution, although less important than in the case of charm.
Its effect in the central $\eta$ region is however marginal.
We are therefore able
to give a reliable prediction of the $p_{\sss T}$ spectrum for
$b$ hadrons in the central $\eta$ region.

Finally, we compared our results with those obtained using the
parton shower Monte Carlo HERWIG. We
limited our discussion to the point-like component for monochromatic
photon-proton collisions. As far as open charm quark is concerned,
we found perfect agreement for the $p_{\sss T}$ distribution,
in the whole range considered. The next-to-leading order
QCD prediction for pseudorapidity distribution is instead
slightly broader than the one given by HERWIG in the region
around the peak. This difference is however immaterial in the
central pseudorapidity region, which is explored by the experiments.
We also point out that, without any small transverse momentum cut,
differential cross sections in ``longitudinal" quantities such as $\eta$
are indeed not characterized by a true hard scale. In this case,
differences between perturbative QCD and parton shower approach
are expected. Coming to charmed hadron production, we compared
HERWIG prediction with QCD calculation supplemented with the Peterson
fragmentation function. A small-$p_{\sss T}$ cut was applied.
The agreement is surprisingly good in the whole $\eta$ range,
while the two predictions for transverse momentum distribution
differ in shape only in the small-$p_{\sss T}$ region.

\section*{Acknowledgements}
We wish to thank G.~Iacobucci, M.~Mangano, J.~Roldan, M.~Seymour and
B.~Webber for useful discussions.

\clearpage
\appendix
\section*{Subtraction schemes in photoproduction}
A differential photoproduction cross section can be written as
\beq
d\sigma^{(\gamma {\sss H})}(P_\gamma,P_{\sss H})\,=\,
d\sigma^{(\gamma {\sss H})}_{\rm point}(P_\gamma,P_{\sss H})+
d\sigma^{(\gamma {\sss H})}_{\rm hadr}(P_\gamma,P_{\sss H})\,,
\label{fullxsec}
\eeq
where the quantities in the right-hand side of this equation are the so-called
point-like (or direct) and hadronic (or resolved) photon cross
sections.
In QCD~[\ref{CSS}],
\beqn
d\sigma_{\rm point}^{(\gamma {\sss H})}(P_\gamma,P_{\sss H})&=&\sum_j\int dx
f^{({\sss H})}_j(x,\muf)
d\hat{\sigma}_{\gamma j}(xP_{\sss H},\as(\mur),\mur,\muf,\mug)
\label{pointcomp}
\\
d\sigma_{\rm hadr}^{(\gamma {\sss H})}(P_\gamma,P_{\sss H})
&=&\sum_{ij}\int dx dy
f^{(\gamma)}_i(x,\mug) f^{({\sss H})}_j(y,\muf)
d\hat{\sigma}_{ij}(xP_\gamma,yP_{\sss H},\as(\mur),\mur,\muf,\mug)\,.
\nonumber \\
\label{hadrcomp}
\eeqn
In the case of heavy quark production at next-to-leading order,
the short distance partonic cross sections are given by
\beqn
d\hat{\sigma}_{\gamma j}(p_1,p_2,\as(\mur),\mur,\muf,\mug)&=&
\aem\as(\mur)d\sigma_{\gamma j}^{(0)}(p_1,p_2)
\nonumber \\
&&+\aem\as^2(\mur)d\hat{\sigma}_{\gamma j}^{(1)}(p_1,p_2,\mur,\muf,\mug)
\label{pointxsec}
\\
d\hat{\sigma}_{ij}(p_1,p_2,\as(\mur),\mur,\muf,\mug)&=&
\as^2(\mur)d\sigma_{ij}^{(0)}(p_1,p_2)
\nonumber \\
&&+\as^3(\mur)d\hat{\sigma}_{ij}^{(1)}(p_1,p_2,\mur,\muf,\mug)\,.
\label{hadrxsec}
\eeqn

In dimensional regularization, we have
\beqn
d\hat{\sigma}_{\gamma j}^{(1)}(p_1,p_2)&=&
d\sigma_{\gamma j}^{(1)}(p_1,p_2,\frac{1}{\epb})
+\frac{1}{2\pi}\sum_k\int dx \left(\frac{1}{\epb}P_{k\gamma}(x)
-H_{k\gamma}(x)\right)d\sigma_{kj}^{(0)}(xp_1,p_2)\,
\nonumber \\
&&+\frac{1}{2\pi}\sum_k\int dx \left(\frac{1}{\epb}P_{kj}(x)-
K^{({\sss H})}_{kj}(x)\right) d\sigma_{\gamma k}^{(0)}(p_1,xp_2);
\label{subpoint}
\\
d\hat{\sigma}_{ij}^{(1)}(p_1,p_2)&=&
d\sigma_{ij}^{(1)}(p_1,p_2,\frac{1}{\epb})
+\frac{1}{2\pi}\sum_k\int dx \left(\frac{1}{\epb}P_{ki}(x)-
K_{ki}^{(\gamma)}(x)\right) d\sigma_{kj}^{(0)}(xp_1,p_2)
\nonumber \\
&&+\frac{1}{2\pi}\sum_k\int dx \left(\frac{1}{\epb}P_{kj}(x)-
K_{kj}^{({\sss H})}(x)\right) d\sigma_{ik}^{(0)}(p_1,xp_2)\,,
\label{subhadr}
\eeqn
where $d\sigma_{\gamma j}^{(1)}$ ($d\sigma_{\gamma j}^{(0)}$)
and $d\sigma_{ij}^{(1)}$ ($d\sigma_{ij}^{(0)}$) are the full,
$d$-dimensional regulated partonic cross sections at the next-to-leading
(leading) order for the point-like and hadronic contributions respectively.
The $1/\epb$ singularities in $d\sigma_{ij}^{(1)}$ and
$d\sigma_{\gamma j}^{(1)}$ are appropriately subtracted on the right-hand
side of eqs.~(\ref{subpoint}), (\ref{subhadr}).

The $P_{ij}$ are the usual
Altarelli-Parisi splitting function; $P_{j\gamma}$ is the kernel for
the splitting process $\gamma\to j+\bar{j}$, which is equal to $P_{qg}$
up to a colour factor. The functions $K_{ij}^{({\sss H})}$,
$K_{ij}^{(\gamma)}$ and $H_{k\gamma}$
are completely arbitrary, in that they define an extra finite part of the
subtraction; different choices correspond to different subtraction schemes.
In eqs.~(\ref{subpoint}) and~(\ref{subhadr}) the $\MSB$ scheme is equivalent
to $H=K=0$. For greater generality, we have admitted the possibility to
have different subtraction schemes on photon and hadron legs.
A change in the subtraction scheme implies that parton distribution functions
are modified as follows
\beq
f_i^\prime=f_i+\frac{\aem}{2\pi}H_i+\frac{\as}{2\pi}\sum_{j}K_{ij}\otimes f_j.
\eeq
The term $H$ is present only in the photon case, and it is a direct
consequence of the inhomogeneous term in the modified Altarelli-Parisi
equations. In ref.~[\ref{GRVdis}] a factorization scheme ($\DIG$) for
the photon densities is introduced, which uses $K=0$ and $H\ne 0$.
As can be seen for eqs.~(\ref{subpoint}) and~(\ref{subhadr}), the
$\DIG$ scheme is therefore equivalent to $\MSB$ as far as the hadronic
component is concerned. It does however modify the point-like component through
the second term in the right-hand side of eq.~(\ref{subpoint}), which is
also responsible for a subtraction on the photon leg.
This clearly shows that the point-like and the hadronic
component are closely related, and that they should not be
considered separately. In principle, when considering the variation
with respect to $\mug$, cancellations occur between the hadronic and
the point-like components, resulting in renormalization group
invariance of the physical cross section $d\sigma^{(\gamma H)}$
up to terms of order $\aem\as^3$.

Photoproduction cross sections are usually linked to electroproduction
ones by a convolution with the Weizs\"acker-Williams function~[\ref{WW}]:
\beq
d\sigma^{(e{\sss H})}(P_e,P_{\sss H})\,=\,\int dx f(x,\mu_{\sss WW})
d\sigma^{(\gamma {\sss H})}(xP_e,P_{\sss H})\,.
\label{etogamma}
\eeq
At HERA, the H1 and ZEUS experiments can directly investigate the
virtuality of the exchanged photon, by tagging the emitted electron
and retaining only those events in which the electron scattering
angle $\theta$ satisfies the condition $\theta <\theta_c$, with
$\theta_c$ typically of the order of few mrad. In ref.~[\ref{FMNRww}]
a functional form especially suited for this experimental set-up
was proposed, namely
\beqn
f(x,E_e)&=&\frac{\aem}{2\pi}\Bigg\{2(1-x)\left[
\frac{\me^2 x}{E_e^2(1-x)^2\theta_c^2+\me^2 x^2}-\frac{1}{x}\right]
\nonumber \\&& \phantom{\frac{\aem}{2\pi}\Bigg\{}
+\frac{1+(1-x)^2}{x}\log\frac{E_e^2(1-x)^2\theta_c^2+\me^2 x^2}
{\me^2 x^2}\Bigg\}\,,
\eeqn
where $E_e$ is the incoming electron energy in the laboratory frame.
When no angular cut is applied, eq.~(\ref{etogamma}) is too restrictive;
we relax it by writing
\beq
d\sigma^{(e{\sss H})}(P_e,P_{\sss H})\,=\,\int dx f(x,\mu_{\sss WW})
d\sigma^{(\gamma {\sss H})}_{point}(xP_e,P_{\sss H})+
\int dx f(x,\mu_{\sss WW}^\prime)
d\sigma^{(\gamma {\sss H})}_{hadr}(xP_e,P_{\sss H})\,,
\label{etogamma2}
\eeq
with $\mu_{\sss WW}\neq\mu_{\sss WW}^\prime$ (for a discussion
of the $\mu_{\sss WW}$ and $\mu_{\sss WW}^\prime$ values,
see ref.~[\ref{FMNR_TOT}] and references therein).
Strictly speaking, the fact that $\mu_{\sss WW}\neq\mu_{\sss WW}^\prime$
gives up the renormalization group invariance of the physical cross section
in eq.~(\ref{etogamma2}), in that no cancellation occurs between the
point-like and the hadronic component when variation with respect to
$\mug$ is considered. In practice, we have verified that this effect
is numerically negligible, the point-like component being almost
invariant with respect to $\mug$ variations.

\clearpage
\begin{reflist}
\item \label{ZeusCharm}
    M.~Derrick {\it et al.}, ZEUS Coll., preprint DESY-95-013, 1995.
\item \label{FixedTargetPhotoproduction}
    P.~Frabetti {\it et al.}, E687 Coll., \pl{B308}{93}{193};\newline
    J.~C.~Anjos {\it et al.}, E691 Coll., \prl{62}{89}{513};\newline
    M.~P.~Alvarez {\it et al.}, NA14/2 Coll., \zp{C60}{93}{53};\newline
    G.~Bellini, Proceedings of ``Les Rencontres de Physique de la Vall\'ee
    d'Aoste'', La Thuile, Aosta Valley, 6-12 March, 1994.
\item \label{FMNR_TOT}
    S.~Frixione, M.L.~Mangano, P.~Nason and G.~Ridolfi,
    \pl{B348}{95}{633}.
\item\label{EllisNason}
    R.K.~Ellis and P.~Nason, \np{B312}{89}{551}.
\item \label{NDE}
   P.~Nason, S.~Dawson and R.K.~Ellis, \np{B303}{88}{607}; {\bf B327}(1988)49.
\item \label{MNR}
   M.L.~Mangano, P.~Nason and G.~Ridolfi, \np{B373}{92}{295}.
\item \label{FMNRphoto}
   S.~Frixione, M.L.~Mangano, P.~Nason and G.~Ridolfi, \np{B412}{94}{225}.
\item \label{VirtualPhoton}
   E.~Laenen, S.~Riemersma, J.~Smith and W.L.~van~Neerven, \\
   \np{B392}{93}{162};\\
   S.~Riemersma, J.~Smith and W.L.~van~Neerven, \pl{B347}{95}{143};\\
   B.W.~Harris and J.~Smith, preprint ITP-SB-95-08, hep-ph/9503484.
\item\label{HERWIG}
    G.~Marchesini, B.R.~Webber, G.~Abbiendi, I.~G.~Knowles, \\
    M.~H.~Seymour and L.~ Stanco, {\it Comp. Phys. Comm.}~{\bf 67}(1992)465.
\item\label{MRSA}
   A.D.~Martin, R.G.~Roberts and W.J.~Stirling, \pr{D50}{94}{6734}.
\item\label{MRSG}
   A.D.~Martin, R.G.~Roberts and W.J.~Stirling, preprint RAL-95-021,
   DTP/95/14, hep-ph/9502336.
\item \label{pete}
   C.~Peterson, D.~Schlatter, I.~Schmitt and P.~Zerwas, \pr{D27}{83}{105}.
\item\label{Chrin}
   J.~Chrin, \zp{C36}{87}{163}.
\item \label{FMNRft}
    S.~Frixione, M.L.~Mangano, P.~Nason and G.~Ridolfi,
    \np{B431}{94}{453}.
\item\label{MRSDmeno}
   A.D.~Martin, R.G.~Roberts and~W.J.~Stirling, \pl{B306}{93}{145},
   \pl{B309}{93}{492}.
\item\label{CTEQ2MF}
   H.L.~Lai {\it et al.}, preprint MSU-HEP-41024, CTEQ-404, hep-ph/9410404.
\item\label{GRV_lam}
   A.~Vogt, preprint DESY 95-068, hep-ph/9504285.
\item \label{LAC1}
   H.~Abramowicz, K.~Charchula and A.~Levy, \pl{269B}{91}{458}.
\item \label{GRVph}
    M.~Gl\"uck, E.~Reya and A.~Vogt, \pr{D46}{92}{1973}.
\item \label{FMNRww}
   S.~Frixione, M.L.~Mangano, P.~Nason and G.~Ridolfi,
   \pl{B319}{93}{339}.
\item \label{CSS}
    J.C.~Collins, D.E.~Soper and G.~Sterman, in {\it Perturbative Quantum\\
    Chromodynamics}, 1989, ed. A.H. Mueller, World Scientific, Singapore, and
    references therein.
\item \label{GRVdis}
    M.~Gl\"uck, E.~Reya and A.~Vogt, \pr{D45}{92}{3986}.
\item \label{WW}
    C.F.~Weizs\"acker, \zp{88}{34}{612};\\
    E.J.~Williams, \pr{45}{34}{729}.
\end{reflist}
\end{document}